\def\DIRvalue{Dimofte}
\def\IDvalue{DI}
\def\titlevalue{Perturbative and nonperturbative aspects of complex Chern-Simons Theory}
\def\authorvalue{Tudor Dimofte}
\def\shortauthorvalue{\authorvalue}
\def\addressvalue{Perimeter Institute for Theoretical Physics, 31 Caroline St. N, Waterloo, ON N2J 2Y5, Canada\\
(on leave) Department of Mathematics, University of California, Davis, CA 95616, USA\\
  \tt tudor@math.ucdavis.edu}

\def\abstractvalue{We present an elementary review of some aspects of Chern-Simons theory with complex gauge group $SL(N,\mathbb C)$. We discuss some of the challenges in defining the theory as a full-fledged TQFT, as well as some successes inspired by the 3d-3d correspondence. The 3d-3d correspondence relates partition functions (and other aspects) of complex Chern-Simons theory on a 3-manifold $M$ to supersymmetric partition functions (and other observables) in an associated 3d theory $T[M]$. Many of these observables may be computed by supersymmetric localization. We present several prominent applications to 3-manifold topology and number theory in light of the 3d-3d correspondence.}

\def\preprintvalue{}

\ifx\ifLONG\undefined
\documentclass[12pt]{article}
\newcommand{\chapterauthor}[1]{
\begin{center}
{\bf \normalsize  #1}
\end{center}
}

\newcommand{\chapteraddress}[1]{
\begin{center}
{ \small \it \addressvalue}
\end{center}
}

\newcommand{\chapterabstract}[1]{
\vspace{\baselineskip}
\begin{center}
\textbf{\small Abstract}
\end{center}
#1}

\newcommand{\chapterheader}{

\chapter[\titlevalue{}  (by \shortauthorvalue)]{\titlevalue}
\label{Chapter\IDvalue}
\chapterauthor{\authorvalue}
\chapteraddress{\addressvalue}
\chapterabstract{\abstractvalue}
\tightmtctrue
\minitoc
}

\newcommand{\documentheader}{
\begin{flushright} \small
  \preprintvalue
 \end{flushright}

\begin{center}
{\bf \Large \titlevalue}
\end{center}

\chapterauthor{\authorvalue}
\chapteraddress{\addressvalue}
\chapterabstract{\abstractvalue}

\medskip

This is a contribution to the review volume ``Localization techniques
in quantum field theories'' (eds. V.~Pestun and M.~Zabzine) which
contains 17 Chapters available at \cite{ContributionSummary}

\tableofcontents
}

\newcommand{\ifvolume}[2]{\ifx\ifLONG\undefined#2\else#1\fi}

\newcommand{\documentfinish}{
\ifx\ifLONG\undefined
\bibliographystyle{bibreview} 
\bibliography{\IDvalue,review}  
\end{document}
\else
\addcontentsline{toc}{section}{References}
\providecommand{\href}[2]{#2}\begingroup\raggedright\begin{thebibliography}{100}

\bibitem{ContributionSummary}
V.~Pestun and M.~Zabzine, eds., {\em Localization techniques in quantum field
  theory}, vol.~xx.
\newblock Journal of Physics A, 2016.
\newblock \href{http://arxiv.org/abs/1608.02952}{{\tt 1608.02952}}.
\newblock \url{https://arxiv.org/src/1608.02952/anc/LocQFT.pdf},
  \url{http://pestun.ihes.fr/pages/LocalizationReview/LocQFT.pdf}.

\bibitem{DIAT-CSgrav}
A.~Achucarro and P.~K. Townsend, ``A Chern-Simons action for three-dimensional
  anti-de Sitter supergravity theories,'' {\em Phys. Lett.} {\bf B180} (1986)
  89--92.

\bibitem{DIWitten-gravCS}
E.~Witten, ``2+1 Dimensional Gravity as an Exactly Soluble System,'' {\em Nucl.
  Phys.} {\bf B311} (1988) no.~1, 46--78.
  \url{http://www.ams.org/mathscinet/search/publications.html?pg1=MR&s1=MR974271}.

\bibitem{DIWitten-top}
E.~Witten, ``Topology-Changing Amplitudes in (2+1)-Dimensional Gravity,'' {\em
  Nucl. Phys.} {\bf B323} (1989)  113.
  \url{http://adsabs.harvard.edu/abs/1989NuPhB.323..113W}.

\bibitem{DIWitten-cx}
E.~Witten, ``Quantization of Chern-Simons Gauge Theory with Complex Gauge
  Group,'' {\em Comm. Math. Phys} {\bf 137} (1991)  29--66.

\bibitem{DICFPT-spin3}
A.~Campoleoni, S.~Fredenhagen, S.~Pfenninger, and S.~Theisen, ``Asymptotic
  symmetries of three-dimensional gravity coupled to higher-spin fields,''
  \href{http://dx.doi.org/10.1007/JHEP11(2010)007}{{\em JHEP} {\bf 1011} (2010)
   007}, \href{http://arxiv.org/abs/1008.4744v2}{{\tt 1008.4744v2}}.
  \url{http://arxiv.org/abs/1008.4744v2}.

\bibitem{DIGaberdielGopakumar}
M.~R. Gaberdiel and R.~Gopakumar, ``Minimal Model Holography,''
  \href{http://dx.doi.org/10.1088/1751-8113/46/21/214002}{{\em J. Phys.} {\bf
  A46} (2013)  214002}, \href{http://arxiv.org/abs/1207.6697v2}{{\tt
  1207.6697v2}}. \url{http://arxiv.org/abs/1207.6697v2}.

\bibitem{DICJ-S3}
C.~Cordova and D.~L. Jafferis, ``Complex Chern-Simons from M5-branes on the
  Squashed Three-Sphere,'' \href{http://arxiv.org/abs/1305.2891v1}{{\tt
  1305.2891v1}}. \url{http://arxiv.org/abs/1305.2891v1}.

\bibitem{DILY-S2}
S.~Lee and M.~Yamazaki, ``3d Chern-Simons Theory from M5-branes,'' {\em JHEP}
  {\bf 1312} (2013)  035, \href{http://arxiv.org/abs/1305.2429v2}{{\tt
  1305.2429v2}}. \url{http://arxiv.org/abs/1305.2429v2}.

\bibitem{DIYagi-S2}
J.~Yagi, ``3d TQFT from 6d SCFT,'' {\em JHEP} {\bf 1308} (2013)  017,
  \href{http://arxiv.org/abs/1305.0291v1}{{\tt 1305.0291v1}}.
  \url{http://arxiv.org/abs/1305.0291v1}.

\bibitem{DIDGH}
T.~Dimofte, S.~Gukov, and L.~Hollands, ``Vortex Counting and Lagrangian
  3-manifolds,'' {\em Lett. Math. Phys.} {\bf 98} (2011)  225--287,
  \href{http://arxiv.org/abs/1006.0977v1}{{\tt 1006.0977v1}}.
  \url{http://arxiv.org/abs/1006.0977v1}.

\bibitem{DIDG-Sdual}
T.~Dimofte and S.~Gukov, ``Chern-Simons Theory and S-duality,'' {\em JHEP} {\bf
  1305} (2013)  109, \href{http://arxiv.org/abs/1106.4550v1}{{\tt
  1106.4550v1}}. \url{http://arxiv.org/abs/1106.4550v1}.

\bibitem{DIYamazaki-3d}
Y.~Terashima and M.~Yamazaki, ``SL(2,R) Chern-Simons, Liouville, and Gauge
  Theory on Duality Walls,'' {\em JHEP} {\bf 1108} (2011)  135,
  \href{http://arxiv.org/abs/1103.5748v1}{{\tt 1103.5748v1}}.
  \url{http://arxiv.org/abs/1103.5748v1}.

\bibitem{DIDGG}
T.~Dimofte, D.~Gaiotto, and S.~Gukov, ``Gauge Theories Labelled by
  Three-Manifolds,'' {\em Comm. Math. Phys.} {\bf 325} (2014)  367--419,
  \href{http://arxiv.org/abs/1108.4389v1}{{\tt 1108.4389v1}}.
  \url{http://arxiv.org/abs/1108.4389v1}.

\bibitem{DIAGT}
L.~F. Alday, D.~Gaiotto, and Y.~Tachikawa, ``Liouville Correlation Functions
  from Four-Dimensional Gauge Theories,'' {\em Lett. Math. Phys.} {\bf 91}
  (2010) no.~2, 167--197, \href{http://arxiv.org/abs/0906.3219v1}{{\tt
  0906.3219v1}}. \url{http://arxiv.org/abs/0906.3219v1}.

\bibitem{DIGGP-4d}
A.~Gadde, S.~Gukov, and P.~Putrov, ``Fivebranes and 4-manifolds,''
  \href{http://arxiv.org/abs/1306.4320v1}{{\tt 1306.4320v1}}.
  \url{http://arxiv.org/abs/1306.4320v1}.

\bibitem{DIWit-Jones}
E.~Witten, ``Quantum Field Theory and the Jones Polynomial,'' {\em Comm. Math.
  Phys.} {\bf 121} (1989) no.~3, 351--399.

\bibitem{DIResh-Tur}
N.~Reshetikhin and V.~Turaev, ``Invariants of 3-manifolds via link polynomials
  and quantum groups,'' {\em Invent. Math.} {\bf 103} (1991) no.~3, 547--597.
  \url{http://www.springerlink.com/index/U416971M947560R7.pdf}.

\bibitem{DIKashAnd}
J.~E. Andersen and R.~Kashaev, ``A TQFT from quantum Teichm{\"u}ller theory,''
  \href{http://arxiv.org/abs/1109.6295v1}{{\tt 1109.6295v1}}.
  \url{http://arxiv.org/abs/1109.6295v1}.

\bibitem{DIDGG-index}
T.~Dimofte, D.~Gaiotto, and S.~Gukov, ``3-Manifolds and 3d Indices,'' {\em Adv.
  Theor. Math. Phys.} {\bf 17} (2013)  975--1076,
  \href{http://arxiv.org/abs/1112.5179v1}{{\tt 1112.5179v1}}.
  \url{http://arxiv.org/abs/1112.5179v1}.

\bibitem{DIAK-complexCS}
J.~E. Andersen and R.~Kashaev, ``Complex Quantum Chern-Simons,''
  \href{http://arxiv.org/abs/1409.1208v1}{{\tt 1409.1208v1}}.
  \url{http://arxiv.org/abs/1409.1208v1}.

\bibitem{DID-levelk}
T.~Dimofte, ``Complex Chern-Simons theory at level k via the 3d-3d
  correspondence,'' {\em Comm. Math. Phys.} {\bf 339} (2015) no.~2, 619--662,
  \href{http://arxiv.org/abs/1409.0857v1}{{\tt 1409.0857v1}}.
  \url{http://arxiv.org/abs/1409.0857v1}.

\bibitem{DIGukovPei}
S.~Gukov and D.~Pei, ``Equivariant Verlinde formula from fivebranes and
  vortices,'' \href{http://arxiv.org/abs/1501.01310v1}{{\tt 1501.01310v1}}.
  \url{http://arxiv.org/abs/1501.01310v1}.

\bibitem{DIPeiKe}
D.~Pei and K.~Ye, ``A 3d-3d appetizer,''
  \href{http://arxiv.org/abs/1503.04809v1}{{\tt 1503.04809v1}}.
  \url{http://arxiv.org/abs/1503.04809v1}.

\bibitem{DIgukov-2003}
S.~Gukov, ``Three-Dimensional Quantum Gravity, Chern-Simons Theory, and the
  A-Polynomial,'' {\em Commun. Math. Phys.} {\bf 255} (2005) no.~3, 577--627,
  \href{http://arxiv.org/abs/hep-th/0306165v1}{{\tt hep-th/0306165v1}}.
  \url{http://arxiv.org/abs/hep-th/0306165v1}.

\bibitem{DIkashaev-1997}
R.~M. Kashaev, ``The hyperbolic volume of knots from quantum dilogarithm,''
  {\em Lett. Math. Phys.} {\bf 39} (1997)  269--265,
  \href{http://arxiv.org/abs/q-alg/9601025v2}{{\tt q-alg/9601025v2}}.
  \url{http://arxiv.org/abs/q-alg/9601025v2}.

\bibitem{DIMur-Mur}
H.~Murakami and J.~Murakami, ``The colored Jones polynomials and the simplicial
  volume of a knot,'' {\em Acta Math.} {\bf 186} (2001) no.~1, 85--104,
  \href{http://arxiv.org/abs/math/9905075v2}{{\tt math/9905075v2}}.
  \url{http://arxiv.org/abs/math/9905075v2}.

\bibitem{DIWit-anal}
E.~Witten, ``Analytic Continuation of Chern-Simons Theory,'' {\em Chern-Simons
  gauge theory: 20 Years After (AMS/IP Stud. Adv. Math.)} (2011)  347--446,
  \href{http://arxiv.org/abs/1001.2933v2}{{\tt 1001.2933v2}}.
  \url{http://arxiv.org/abs/1001.2933v4}.

\bibitem{DIDF}
R.~Dijkgraaf and H.~Fuji, ``The Volume Conjecture and Topological Strings,''
  {\em Fortschr. Phys.} {\bf 57} (2009) no.~9, 825--856,
  \href{http://arxiv.org/abs/0903.2084v2}{{\tt 0903.2084v2}}.
  \url{http://arxiv.org/abs/0903.2084v2}.

\bibitem{DIGS-quant}
S.~Gukov and P.~Su{\l}kowski, ``A-polynomial, B-model, and Quantization,'' {\em
  JHEP} {\bf 1202} (2012)  070, \href{http://arxiv.org/abs/1108.0002v1}{{\tt
  1108.0002v1}}. \url{http://arxiv.org/abs/1108.0002v1}.

\bibitem{DIBorotEynard}
G.~Borot and B.~Eynard, ``All-order asymptotics of hyperbolic knot invariants
  from non-perturbative topological recursion of A-polynomials,''
  \href{http://arxiv.org/abs/1205.2261v2}{{\tt 1205.2261v2}}.
  \url{http://arxiv.org/abs/1205.2261v2}.

\bibitem{DIEO}
B.~Eynard and N.~Orantin, ``Invariants of Algebraic Curves and Topological
  Expansion,'' {\em Commun. Number Theory Phys.} {\bf 1} (2007) no.~2,
  347--452, \href{http://arxiv.org/abs/math-ph/0702045v4}{{\tt
  math-ph/0702045v4}}. \url{http://arxiv.org/abs/math-ph/0702045v4}.

\bibitem{DIFGS-VC}
H.~Fuji, S.~Gukov, and P.~Su{\l}kowski, ``Volume Conjecture: Refined and
  Categorified,'' \href{http://arxiv.org/abs/1203.2182v1}{{\tt 1203.2182v1}}.
  \url{http://arxiv.org/abs/1203.2182v1}.

\bibitem{DIGPV}
S.~Gukov, P.~Putrov, and C.~Vafa, ``Fivebranes and 3-manifold homology,''
  \href{http://arxiv.org/abs/1602.05302v1}{{\tt 1602.05302v1}}.
  \url{http://arxiv.org/abs/1602.05302v1}.

\bibitem{DIDG-levelk}
T.~Dimofte and S.~Garoufalidis, ``Quantum modularity and complex Chern-Simons
  theory,'' \href{http://arxiv.org/abs/1511.05628v1}{{\tt 1511.05628v1}}.
  \url{http://arxiv.org/abs/1511.05628v1}.

\bibitem{DIQMF}
D.~Zagier, ``Quantum Modular Forms,'' {\em Quanta of Maths, Clay Math. Proc.}
  {\bf 11} (2010)  659--675. AMS, Prividence, RI.

\bibitem{DITY-cluster}
Y.~Terashima and M.~Yamazaki, ``3d N=2 Theories from Cluster Algebras,'' {\em
  PTEP} {\bf 023} (2014)  B01, \href{http://arxiv.org/abs/1301.5902v1}{{\tt
  1301.5902v1}}. \url{http://arxiv.org/abs/1301.5902v1}.

\bibitem{DIDGG-Kdec}
T.~Dimofte, M.~Gabella, and A.~B. Goncharov, ``K-Decompositions and 3d Gauge
  Theories,'' \href{http://arxiv.org/abs/1301.0192v1}{{\tt 1301.0192v1}}.
  \url{http://arxiv.org/abs/1301.0192v1}.

\bibitem{DIFockChekhov}
L.~Chekhov and V.~V. Fock, ``Quantum Teichm{\"u}ller Space,'' {\em Theoret. and
  Math. Phys.} {\bf 120} (1999) no.~3, 1245--1259,
  \href{http://arxiv.org/abs/math/9908165v2}{{\tt math/9908165v2}}.
  \url{http://arxiv.org/abs/math/9908165v2}.

\bibitem{DIKash-Teich}
R.~M. Kashaev, ``Quantization of Teichm{\"u}ller Spaces and the Quantum
  Dilogarithm,'' {\em Lett. Math. Phys.} {\bf 43} (1998) no.~2, 105--115.
  \url{http://www.ams.org/mathscinet/search/publications.html?pg1=MR&s1=MR1607296}.

\bibitem{DIFG-cluster}
V.~V. Fock and A.~B. Goncharov, ``Cluster ensembles, quantization and the
  dilogarithm,'' {\em Annales Scientifiques L'Ecole Normal Superier, 4-e
  series} {\bf t.42} (2009)  865--930,
  \href{http://arxiv.org/abs/math/0311245v5}{{\tt math/0311245v5}}.
  \url{http://arxiv.org/abs/math/0311245v5}.

\bibitem{DIVafa-FQHE}
C.~Vafa, ``Fractional Quantum Hall Effect and M-Theory,''
  \href{http://arxiv.org/abs/1511.03372v3}{{\tt 1511.03372v3}}.
  \url{http://arxiv.org/abs/1511.03372v3}.

\bibitem{DIatiyah-1990}
M.~Atiyah, ``Geometry and Physics of Knots,'' {\em Cambridge Univ. Press,
  Cambridge} (1990)  78pp.
  \url{http://books.google.com/books?hl=en&lr=&id=lOWaaHA5rBcC&oi=fnd&pg=PA291&dq=%2522atiyah%2522+%2522knots%2522&ots=55BgFRpwVT&sig=1Rg8AWBVtC1ibLt7986jgkmK4Vw}.

\bibitem{DICDGS}
H.-J. Chung, T.~Dimofte, S.~Gukov, and P.~Su{\l}kowski, ``3d-3d Correspondence
  Revisited,'' \href{http://arxiv.org/abs/1405.3663v1}{{\tt 1405.3663v1}}.
  \url{http://arxiv.org/abs/1405.3663v1}.

\bibitem{DIGHHR}
S.~Garoufalidis, C.~Hodgson, N.~Hoffman, and H.~Rubinstein, ``{The 3D-index and
  normal surfaces},''
\href{http://arxiv.org/abs/1604.02688}{{\tt arXiv:1604.02688 [math.GT]}}.

\bibitem{ContributionDU}
T.~Dumitrescu, ``An Introduction to Supersymmetric Field Theories in Curved
  Space,'' {\em Journal of Physics A} {\bf xx} (2016)  000,
  \href{http://arxiv.org/abs/1608.02957}{{\tt 1608.02957}}.

\bibitem{ContributionWI}
B.~Willett, ``Localization on three-dimensional manifolds,'' {\em Journal of
  Physics A} {\bf xx} (2016)  000, \href{http://arxiv.org/abs/1608.02958}{{\tt
  1608.02958}}.

\bibitem{DIAG-HW}
J.~E. Andersen and N.~L. Gammelgaard, ``The Hitchin-Witten Connection and
  Complex Quantum Chern-Simons Theory,''
  \href{http://arxiv.org/abs/1409.1035v1}{{\tt 1409.1035v1}}.
  \url{http://arxiv.org/abs/1409.1035v1}.

\bibitem{DIDGLZ}
T.~Dimofte, S.~Gukov, J.~Lenells, and D.~Zagier, ``Exact Results for
  Perturbative Chern-Simons Theory with Complex Gauge Group,'' {\em Comm. Num.
  Thy. and Phys.} {\bf 3} (2009) no.~2, 363--443,
  \href{http://arxiv.org/abs/0903.2472v1}{{\tt 0903.2472v1}}.
  \url{http://arxiv.org/abs/0903.2472v1}.

\bibitem{DIGW-branes}
S.~Gukov and E.~Witten, ``Branes and Quantization,'' {\em Adv. Theor. Math.
  Phys.} {\bf 13} (2009) no.~5, 1445--1518,
  \href{http://arxiv.org/abs/0809.0305v2}{{\tt 0809.0305v2}}.
  \url{http://arxiv.org/abs/0809.0305v2}.

\bibitem{DImirrorbranes}
S.~Gukov, ``Quantization via Mirror Symmetry,''
  \href{http://arxiv.org/abs/1011.2218v1}{{\tt 1011.2218v1}}.
  \url{http://arxiv.org/abs/1011.2218v1}.

\bibitem{DIHitchin-SD}
N.~Hitchin, ``The self-duality equations on a Riemann surface,'' {\em Proc.
  London Math. Soc} (1987)  .
  \url{http://plms.oxfordjournals.org/cgi/reprint/s3-55/1/59.pdf}.

\bibitem{DIaxelrod-witten}
S.~Axelrod, S.~D. Pietra, and E.~Witten, ``Geometric Quantization of
  Chern-Simons Gauge Theory,'' {\em J. Diff. Geom.} {\bf 33} (1991) no.~3,
  787--902.
  \url{http://www.ams.org/mathscinet/search/publications.html?pg1=MR&s1=MR1100212}.

\bibitem{DIDGV-FQHE}
T.~Dimofte, S.~Gukov, and C.~Vafa, ``Work in progress,''.

\bibitem{DINT-para}
T.~Nishioka and Y.~Tachikawa, ``Para-Liouville/Toda central charges from
  M5-branes,'' \href{http://dx.doi.org/10.1103/PhysRevD.84.046009}{{\em Phys.
  Rev.} {\bf D84} (2011)  046009}, \href{http://arxiv.org/abs/1106.1172v1}{{\tt
  1106.1172v1}}. \url{http://arxiv.org/abs/1106.1172v1}.

\bibitem{DICJ-AGT}
C.~Cordova and D.~L. Jafferis, ``{Toda Theory From Six Dimensions},''
\href{http://arxiv.org/abs/1605.03997}{{\tt arXiv:1605.03997 [hep-th]}}.

\bibitem{DIBDP-blocks}
C.~Beem, T.~Dimofte, and S.~Pasquetti, ``Holomorphic Blocks in Three
  Dimensions,'' \href{http://arxiv.org/abs/1211.1986v1}{{\tt 1211.1986v1}}.
  \url{http://arxiv.org/abs/1211.1986v1}.

\bibitem{DImostow-1973}
G.~Mostow, ``Strong Rigidity of Locally Symmetric Spaces,'' {\em Ann. Math.
  Studies, Princeton Univ. Press, Princeton, NJ} {\bf 78} (1973)  .
  \url{http://books.google.com/books?hl=en&lr=&id=xT0SFmrFrWoC&oi=fnd&pg=PA3&dq=%2522mostow%2522+%2522strong%2522&ots=kACCabOaTW&sig=igChC-yXFJUQLgjr8FoYi6YU5jc}.

\bibitem{DIthurston-1980}
W.~Thurston, ``The Geometry and Topology of Three-Manifolds,'' {\em Lecture
  notes at Princeton University} (1980)  .
  \url{http://pi.unl.edu/~mbrittenham2/classwk/990s08/public/thurston.notes.pdf/6a.pdf}.

\bibitem{DID-volume}
T.~Dimofte, ``3d Superconformal Theories from Three-Manifolds,'' {\em Math.
  Phys. Stud.} {\bf 9783319187693} (2016)  339--373,
  \href{http://arxiv.org/abs/1412.7129v1}{{\tt 1412.7129v1}}.
  \url{http://arxiv.org/abs/1412.7129v1}. J. Teschner, ed.

\bibitem{DIYamazaki-defects}
D.~Gang, N.~Kim, M.~Romo, and M.~Yamazaki, ``Aspects of Defects in 3d-3d
  Correspondence,'' \href{http://arxiv.org/abs/1510.05011v1}{{\tt
  1510.05011v1}}. \url{http://arxiv.org/abs/1510.05011v1}.

\bibitem{DIGaiotto-dualities}
D.~Gaiotto, ``N=2 dualities,'' {\em JHEP} {\bf 1208} (2012)  034,
  \href{http://arxiv.org/abs/0904.2715v1}{{\tt 0904.2715v1}}.
  \url{http://arxiv.org/abs/0904.2715v1}.

\bibitem{DIGMN}
D.~Gaiotto, G.~W. Moore, and A.~Neitzke, ``Four-dimensional wall-crossing via
  three-dimensional field theory,'' {\em Comm. Math. Phys.} {\bf 299} (2010)
  no.~1, 163--224, \href{http://arxiv.org/abs/0807.4723v1}{{\tt 0807.4723v1}}.
  \url{http://arxiv.org/abs/0807.4723v1}.

\bibitem{DICCV}
S.~Cecotti, C.~Cordova, and C.~Vafa, ``Braids, Walls, and Mirrors,''
  \href{http://arxiv.org/abs/1110.2115v1}{{\tt 1110.2115v1}}.
  \url{http://arxiv.org/abs/1110.2115v1}.

\bibitem{DICordova-tangles}
C.~Cordova, S.~Espahbodi, B.~Haghighat, A.~Rastogi, and C.~Vafa, ``Tangles,
  Generalized Reidemeister Moves, and Three-Dimensional Mirror Symmetry,''
  \href{http://arxiv.org/abs/1211.3730v1}{{\tt 1211.3730v1}}.
  \url{http://arxiv.org/abs/1211.3730v1}.

\bibitem{DIAS-refinedCS}
M.~Aganagic and S.~Shakirov, ``Knot Homology from Refined Chern-Simons
  Theory,'' \href{http://arxiv.org/abs/1105.5117v1}{{\tt 1105.5117v1}}.
  \url{http://arxiv.org/abs/1105.5117v1}.

\bibitem{DIGukovMurakami}
S.~Gukov and H.~Murakami, ``SL(2,C) Chern-Simons theory and the asymptotic
  behavior of the colored Jones polynomial,''
  \href{http://dx.doi.org/10.1007/s11005-008-0282-3}{{\em Lett. Math. Phys.}
  {\bf 86} (2008) no.~2-3, 79--98},
  \href{http://arxiv.org/abs/math/0608324v2}{{\tt math/0608324v2}}.
  \url{http://arxiv.org/abs/math/0608324v2}.

\bibitem{DIRaySinger}
D.~B. Ray and I.~M. Singer, ``R-torsion and the Laplacian on Riemannian
  manifolds,'' {\em Adv. in Math.} {\bf 7} (1971)  145--210.
  \url{http://www.ams.org/mathscinet/search/publications.html?pg1=MR&s1=MR0295381}.

\bibitem{DIBW-cx}
D.~Bar-Natan and E.~Witten, ``Perturbative expansion of Chern-Simons theory
  with non-compact gauge group,'' {\em Comm. Math. Phys.} {\bf 141} (1991)
  no.~2, 423--440.
  \url{http://www.springerlink.com/index/G417376W368T0806.pdf}.

\bibitem{DICheegerSimons}
J.~Cheeger and J.~Simons, ``Differential characters and geometric invariants,''
  {\em Lecture Notes in Math.} {\bf 1167} (1985)  50--80. Springer, Berlin.

\bibitem{DIChernSimons}
S.-S. Chern and J.~Simons, ``Characteristic forms and geometric invariants,''
  {\em Ann. of Math} {\bf 99} (1974) no.~2, 48--69.

\bibitem{DIhikami-2006}
K.~Hikami, ``Generalized Volume Conjecture and the A-Polynomials - the
  Neumann-Zagier Potential Function as a Classical Limit of Quantum
  Invariant,'' \href{http://dx.doi.org/10.1016/j.geomphys.2007.03.008}{{\em J.
  Geom. Phys.} {\bf 57} (2007) no.~9, 1895--1940},
  \href{http://arxiv.org/abs/math/0604094v1}{{\tt math/0604094v1}}.
  \url{http://arxiv.org/abs/math/0604094v1}.

\bibitem{DIDG-quantumNZ}
T.~D. Dimofte and S.~Garoufalidis, ``The quantum content of the gluing
  equations,'' {\em Geom. Topol.} {\bf 17} (2013) no.~3, 1253--1315,
  \href{http://arxiv.org/abs/1202.6268v1}{{\tt 1202.6268v1}}.
  \url{http://arxiv.org/abs/1202.6268v1}.

\bibitem{DIHenningsonSkenderis}
M.~Henningson and K.~Skenderis, ``The Holographic Weyl anomaly,''
  \href{http://dx.doi.org/10.1088/1126-6708/1998/07/023}{{\em JHEP} {\bf 9807}
  (1998)  023}, \href{http://arxiv.org/abs/hep-th/9806087v2}{{\tt
  hep-th/9806087v2}}. \url{http://arxiv.org/abs/hep-th/9806087v2}.

\bibitem{DIHMM-anomalies}
J.~A. Harvey, R.~Minasian, and G.~Moore, ``Non-abelian Tensor-multiplet
  Anomalies,'' \href{http://dx.doi.org/10.1088/1126-6708/1998/09/004}{{\em
  JHEP} {\bf 9809} (1998)  004},
  \href{http://arxiv.org/abs/hep-th/9808060v1}{{\tt hep-th/9808060v1}}.
  \url{http://arxiv.org/abs/hep-th/9808060v1}.

\bibitem{DIGKW-5branes}
J.~P. Gauntlett, N.~Kim, and D.~Waldram, ``M-Fivebranes Wrapped on
  Supersymmetric Cycles,''
  \href{http://dx.doi.org/10.1103/PhysRevD.63.126001}{{\em Phys. Rev.} {\bf
  D63} (2001)  126001}, \href{http://arxiv.org/abs/hep-th/0012195v2}{{\tt
  hep-th/0012195v2}}. \url{http://arxiv.org/abs/hep-th/0012195v2}.

\bibitem{DIGTZ-slN}
S.~Garoufalidis, D.~P. Thurston, and C.~K. Zickert, ``The complex volume of
  SL(n,C)-representations of 3-manifolds,''
  \href{http://arxiv.org/abs/1111.2828v1}{{\tt 1111.2828v1}}.
  \url{http://arxiv.org/abs/1111.2828v1}.

\bibitem{DIMPS-S3b}
D.~Martelli, A.~Passias, and J.~Sparks, ``The gravity dual of supersymmetric
  gauge theories on a squashed three-sphere,'' {\em Nucl. Phys.} {\bf B864}
  (2012)  840--868, \href{http://arxiv.org/abs/1110.6400v3}{{\tt 1110.6400v3}}.
  \url{http://arxiv.org/abs/1110.6400v3}.

\bibitem{DIGKL-largeN}
D.~Gang, N.~Kim, and S.~Lee, ``Holography of 3d-3d correspondence at Large N,''
  \href{http://dx.doi.org/10.1007/JHEP04(2015)091}{{\em JHEP} {\bf 1504} (2015)
   091}, \href{http://arxiv.org/abs/1409.6206v1}{{\tt 1409.6206v1}}.
  \url{http://arxiv.org/abs/1409.6206v1}.

\bibitem{DIPorti-largeN}
P.~Menal-Ferrer and J.~Porti, ``Higher dimensional Reidemeister torsion
  invariants for cusped hyperbolic 3-manifolds,''
  \href{http://dx.doi.org/10.1112/jtopol/jtt024}{{\em J. of Topol.} {\bf 7}
  (2014) no.~1, 69--119}, \href{http://arxiv.org/abs/1110.3718v2}{{\tt
  1110.3718v2}}. \url{http://arxiv.org/abs/1110.3718v2}.

\bibitem{DIGar-index}
S.~Garoufalidis, ``The 3D index of an ideal triangulation and angle
  structures,'' \href{http://arxiv.org/abs/1208.1663v1}{{\tt 1208.1663v1}}.
  \url{http://arxiv.org/abs/1208.1663v1}.

\bibitem{DIGHRS-index}
S.~Garoufalidis, C.~D. Hodgson, J.~H. Rubinstein, and H.~Segerman,
  ``1-efficient triangulations and the index of a cusped hyperbolic
  3-manifold,'' \href{http://arxiv.org/abs/1303.5278v1}{{\tt 1303.5278v1}}.
  \url{http://arxiv.org/abs/1303.5278v1}.

\bibitem{DIDimofte-QRS}
T.~Dimofte, ``Quantum Riemann Surfaces in Chern-Simons Theory,'' {\em Adv.
  Theor. Math. Phys.} {\bf 17} (2013)  479--599,
  \href{http://arxiv.org/abs/1102.4847v1}{{\tt 1102.4847v1}}.
  \url{http://arxiv.org/abs/1102.4847v1}.

\bibitem{DIAK-new}
J.~E. Andersen and R.~Kashaev, ``A new formulation of the Teichm{\"u}ller
  TQFT,'' \href{http://arxiv.org/abs/1305.4291v1}{{\tt 1305.4291v1}}.
  \url{http://arxiv.org/abs/1305.4291v1}.

\bibitem{DIKim-index}
S.~Kim, ``The complete superconformal index for N=6 Chern-Simons theory,''
  \href{http://dx.doi.org/10.1016/j.nuclphysb.2009.06.025}{{\em Nucl. Phys.}
  {\bf B821} (2009)  241--284}, \href{http://arxiv.org/abs/0903.4172v5}{{\tt
  0903.4172v5}}. \url{http://arxiv.org/abs/0903.4172v5}.

\bibitem{DIIY-index}
Y.~Imamura and S.~Yokoyama, ``Index for three dimensional superconformal field
  theories with general R-charge assignments,'' {\em JHEP} {\bf 1104} (2011)
  007, \href{http://arxiv.org/abs/1101.0557v3}{{\tt 1101.0557v3}}.
  \url{http://arxiv.org/abs/1101.0557v3}.

\bibitem{DIKW-index}
A.~Kapustin and B.~Willett, ``Generalized Superconformal Index for Three
  Dimensional Field Theories,'' \href{http://arxiv.org/abs/1106.2484v1}{{\tt
  1106.2484v1}}. \url{http://arxiv.org/abs/1106.2484v1}.

\bibitem{DIBNY-lens}
F.~Benini, T.~Nishioka, and M.~Yamazaki, ``4d Index to 3d Index and 2d TQFT,''
  \href{http://dx.doi.org/10.1103/PhysRevD.86.065015}{{\em Phys. Rev.} {\bf
  D86} (2012)  065015}, \href{http://arxiv.org/abs/1109.0283v2}{{\tt
  1109.0283v2}}. \url{http://arxiv.org/abs/1109.0283v2}.

\bibitem{DIIMY-fact}
Y.~Imamura, H.~Matsuno, and D.~Yokoyama, ``Factorization of S3/Zn partition
  function,'' \href{http://dx.doi.org/10.1103/PhysRevD.89.085003}{{\em Phys.
  Rev.} {\bf D89} (2014)  085003}, \href{http://arxiv.org/abs/1311.2371v3}{{\tt
  1311.2371v3}}. \url{http://arxiv.org/abs/1311.2371v3}.

\bibitem{DINZ}
W.~D. Neumann and D.~Zagier, ``Volumes of hyperbolic three-manifolds,'' {\em
  Topology} {\bf 24} (1985) no.~3, 307--332.

\bibitem{DINeumann-combinatorics}
W.~Neumann, ``Combinatorics of Triangulations and the Chern-Simons Invariant
  for Hyperbolic 3-Manifolds,'' {\em in Topology '90, Ohio State Univ. Math.
  Res. Inst. Publ.} {\bf 1} (1992)  .
  \url{http://www.math.columbia.edu/~neumann/preprints/cspaper.pdf}.

\bibitem{DIGK-Gg}
S.~Garoufalidis and R.~Kashaev, ``From state integrals to q-series,''
  \href{http://arxiv.org/abs/1304.2705v1}{{\tt 1304.2705v1}}.
  \url{http://arxiv.org/abs/1304.2705v1}.

\bibitem{DIfaddeev-1994}
L.~Faddeev, ``Current-Like Variables in Massive and Massless Integrable
  Models,'' {\em in Quantum groups and their applications in physics (Varenna
  1994), Proc. Internat. School Phys. Enrico Fermi, 127, IOS, Amsterdam} (1996)
   117--135. \url{http://arxiv.org/abs/hep-th/9408041}.

\bibitem{DIFad-Kash}
L.~D. Faddeev and R.~M. Kashaev, ``Quantum Dilogarithm,'' {\em Modern Phys.
  Lett.} {\bf A9} (1994) no.~5, 427--434,
  \href{http://arxiv.org/abs/hep-th/9310070v1}{{\tt hep-th/9310070v1}}.
  \url{http://arxiv.org/abs/hep-th/9310070v1}.

\bibitem{DIpetronio-2000}
C.~Petronio and J.~Weeks, ``Partially Flat Ideal Triangulations of Cusped
  Hyperbolic 3--Manifolds,'' {\em Osaka J. Math.} {\bf 37} (2000) no.~2,
  453--466.
  \url{http://www.projecteuclid.org/DPubS/Repository/1.0/Disseminate?handle=euclid.ojm/1200789209&view=body&content-type=pdf_1}.

\bibitem{DISnapPy}
M.~Culler, N.~Dunfield, and J.~R. Weeks, ``SnapPy, a computer program for
  studying the geometry and topology of 3-manifolds,''.
  \url{http://snappy.computop.org}. http://snappy.computop.org.

\bibitem{DIWitten-constraints}
E.~Witten, ``Constraints on supersymmetry breaking,'' {\em Nucl. Phys.} {\bf
  B202} (1982) no.~2, 253--316.
  \url{http://www.ams.org/mathscinet/search/publications.html?pg1=MR&s1=MR668987}.

\bibitem{DIGGP-walls}
A.~Gadde, S.~Gukov, and P.~Putrov, ``Walls, Lines, and Spectral Dualities in 3d
  Gauge Theories,'' \href{http://arxiv.org/abs/1302.0015v1}{{\tt 1302.0015v1}}.
  \url{http://arxiv.org/abs/1302.0015v1}.

\bibitem{DIGaroufalidis-slopes}
S.~Garoufalidis, ``The Jones slopes of a knot,'' {\em Quant. Topol.} {\bf 2}
  (2011)  43--69, \href{http://arxiv.org/abs/0911.3627v4}{{\tt 0911.3627v4}}.
  \url{http://arxiv.org/abs/0911.3627v4}.

\bibitem{DIgaroufalidis-2004}
S.~Garoufalidis, ``On the Characteristic and Deformation Varieties of a Knot,''
  {\em Geom. Topol. Monogr.} {\bf 7} (2004)  291--304,
  \href{http://arxiv.org/abs/math/0306230v4}{{\tt math/0306230v4}}.
  \url{http://arxiv.org/abs/math/0306230v4}.

\bibitem{DIcooper-1994}
D.~Cooper, M.~Culler, H.~Gillet, D.~Long, and P.~Shalen, ``Plane Curves
  Associated to Character Varieties of 3-Manifolds,'' {\em Invent. Math.} {\bf
  118} (1994) no.~1, 47--84.
  \url{http://www.springerlink.com/index/R865T605G509312U.pdf}.

\bibitem{DIHatcher-bdy}
A.~Hatcher, ``On the boundary curves of incompressible surfaces,'' {\em Pacific
  J. Math.} {\bf 99} (1982)  373--377.

\bibitem{DIRubinstein-S3}
J.~H. Rubinstein, ``An algorithm to recognize the 3-sphere,'' {\em Proc.
  International Congress of Math.} {\bf 1} (1994) no.~2, 601--611. Birkhauser,
  Basel.

\bibitem{DIJones}
V.~F.~R. Jones, ``A polynomial invariant for knots via von Neumann algebras,''
  {\em Bull. Amer. Math. Soc. (N.S.)} {\bf 12} (1985) no.~1, 103--111.
  \url{http://www.ams.org/mathscinet/search/publications.html?pg1=MR&s1=MR766964}.

\bibitem{DITuraev-YB}
V.~G. Turaev, ``The Yang-Baxter equation and invariants of links,'' {\em
  Invent. Math.} {\bf 92} (1988) no.~3, 527--553.
  \url{http://www.ams.org/mathscinet/search/publications.html?pg1=MR&s1=MR939474}.

\bibitem{DIGK-rational}
S.~Garoufalidis and R.~Kashaev, ``Evaluation of state integrals at rational
  points,'' {\em Comm. Num. Thy. Phys.} {\bf 9} (2015) no.~3, 549--582,
  \href{http://arxiv.org/abs/1411.6062v2}{{\tt 1411.6062v2}}.
  \url{http://arxiv.org/abs/1411.6062v2}.

\end{thebibliography}\endgroup

\fi
}

\newcommand{\documentfinishBBL}{
\addcontentsline{toc}{section}{References}
\ifx\ifLONG\undefined
\input{\IDvalue.separate.bbl}
\end{document}
\else
\input{\DIRvalue/\IDvalue.volume.bbl}
\fi
}

\ifx\ifLONG\undefined
\def\volcite#1{Contribution \cite{Contribution#1}}
\else 
\def\volcite#1{Chapter \ref{Chapter#1}}
\fi

\usepackage{geometry}
\usepackage{amsmath,amssymb,amsfonts,bm}
\usepackage{graphicx,wrapfig}
\usepackage{multirow}
\usepackage{verbatim}
\usepackage{appendix}
\usepackage{rotating}
\usepackage{hyperref}
\usepackage{fullpage}

\numberwithin{equation}{section}

\begin{document}
\thispagestyle{empty}
\documentheader
\newcommand{\CalA}{\mathcal{A}}
\newcommand{\CalB}{\mathcal{B}}
\newcommand{\CalH}{\mathcal{H}}
\newcommand{\CalJ}{\mathcal{J}}
\newcommand{\CalL}{\mathcal{L}}
\newcommand{\CalM}{\mathcal{M}}
\newcommand{\CalN}{\mathcal{N}}
\newcommand{\CalO}{\mathcal{O}}
\newcommand{\CalP}{\mathcal{P}}
\newcommand{\CalS}{\mathcal{S}}
\newcommand{\CalZ}{\mathcal{Z}}

\newcommand{\BZ}{{\mathbb Z}}
\newcommand{\BR}{{\mathbb R}}
\newcommand{\BC}{{\mathbb C}}
\newcommand{\BQ}{{\mathbb Q}}

\newcommand{\Tr}{{\rm Tr \,}}

\else 
\chapterheader 
\fi

\newcommand{\DIbe}{\begin{equation}}
\newcommand{\DIee}{\end{equation}}
\newcommand{\mb}{\mathbf}

\newcommand{\wh}{\widehat}
\newcommand{\ol}{\overline}
\newcommand{\ds}{\displaystyle}
\newcommand{\DIeg}{\emph{e.g.}}
\newcommand{\DIie}{\emph{i.e.}}
\newcommand{\DIcf}{\emph{cf.}}
\newcommand{\cp}{{\mathbb{CP}}}

\newcommand{\bs}{\backslash}
\newcommand{\pd}{\partial}

\numberwithin{equation}{section}
\numberwithin{table}{section}
\numberwithin{figure}{section}

\hyphenation{five-brane space-time}

\section{Introduction}

Chern-Simons theory with complex gauge group came to prominence in the late 80's, partly as a tool for understanding three-dimensional gravity with a negative cosmological constant \cite{DIAT-CSgrav, DIWitten-gravCS, DIWitten-top, DIWitten-cx}.
Many early developments were due to Witten.
Since then, it has found a multitude of applications and deep connections with many parts of theoretical physics and mathematics. A highly incomplete list includes:

\begin{itemize}
\item Many further applications of $SL(2,\BC)$ Chern-Simons theory (and its $SL(2,\BR)\times SL(2,\BR)$ cousin) to three-dimensional quantum gravity and AdS/CFT.
Similarly, $SL(N,\BC)$ Chern-Simons at large $N$ has been used to describe higher-spin theories of gravity~\cite{DICFPT-spin3, DIGaberdielGopakumar}.

\item Chern-Simons theory with gauge group $SL(N,\BC)$ can naturally be embedded in string/M-theory, opening up many powerful perspectives and techniques for analyzing the former.
As a notable example, the compactification of $N$ M5 branes on the product of an ellipsoidally deformed lens space $L(k,1)_b\simeq S^3_b/\BZ_k$ and a three-manifold $M$ (with a topological twist along $M$) leads equivalently to $SL(N,\BC)$ Chern-Simons theory at level $k$ on $M$ \cite{DICJ-S3, DILY-S2, DIYagi-S2} or an $\CalN=2$ supersymmetric theory $T_N[M]$ on the lens space \cite{DIDGH, DIDG-Sdual, DIYamazaki-3d, DIDGG}. This duality, known as the 3d-3d correspondence, fits into a series of dualities involving the compactification of five-branes on various $d$-dimensional manifolds $M^d$, including the AGT correspondence \cite{DIAGT} and the duality of Gukov-Gadde-Putrov relating Vafa-Witten partition functions on $M^4$ and elliptic genera \cite{DIGGP-4d}.

\item There is a multitude of applications to three-dimensional geometry and topology. Fundamentally, partition functions of complex Chern-Simons theory on three-manifolds $M$ provide new topological invariants, generalizing the famous invariants (including knot polynomials) associated with compact Chern-Simons theory \cite{DIWit-Jones, DIResh-Tur}. As yet, a systematic computation of the complex invariants only exists for certain classes of manifolds (\DIeg\ hyperbolic ones \cite{DIKashAnd, DIDGG-index, DIAK-complexCS, DID-levelk}), though new tools to attack the general case are under development \cite{DIGukovPei, DIPeiKe}.

The perturbative expansion of $SL(2,\BC)$ Chern-Simons theory on a three-manifold $M$ encodes various topological invariants of $M$, such as its hyperbolic volume and twisted analytic torsion. In the case that $M=S^3\backslash K$ is a knot complement in $S^3$, it was conjectured by Gukov \cite{DIgukov-2003} that this expansion agrees with a (highly nontrivial) asymptotic limit of colored Jones polynomials of $K$, providing physical motivation for a mathematical statement known as the Volume Conjecture \cite{DIkashaev-1997, DIMur-Mur}. The relation between complex Chern-Simons theory and knot polynomials is essentially a result of analytic continuation, albeit a subtle one \cite{DIWit-anal}.

The perturbative expansion of $SL(2,\BC)$ Chern-Simons theory on knot complements has been successfully reproduced  \cite{DIDF, DIGS-quant, DIBorotEynard} using the topological recursion of Eynard-Orantin \cite{DIEO}, a far-reaching formalism for the quantization of spectral curves.

The study of five-brane systems related to complex Chern-Simons theory recently led to a vast generalization of the Volume Conjecture, involving asymptotic limits of colored HOMFLY polynomials and their categorification \cite{DIFGS-VC}.

\item There are several hints that complex Chern-Simons theory has quasi-modular properties, \DIcf\ \cite{DIDG-Sdual, DIGPV}, though a complete physical characterization of these properties is still missing. The asymptotic expansions of $SL(2,\BC)$ Chern-Simons theory around various singular points in the space of coupling constants (levels) \cite{DIDG-levelk},
related by an action of the modular group, provide evidence for the Quantum Modularity Conjecture of Zagier \cite{DIQMF}.

\item There are close connections between complex Chern-Simons theory and the mathematical theory of cluster algebras, \DIcf\ \cite{DITY-cluster, DIDGG-Kdec}. Cluster algebras play an essential role in the (local) description and quantization of phase spaces that complex Chern-Simons theory attaches to two-dimensional boundaries, \DIcf\ \cite{DIFockChekhov, DIKash-Teich, DIFG-cluster}, and Chern-Simons theory on three-manifolds is associated with cluster-algebra morphisms.

\item Very recently, $SL(2,\BC)$ Chern-Simons theory at integer levels (in terms of \eqref{DI.action} below, this means $k\in \BZ$ and $is\in \BZ$)  has been proposed as an effective theory of quantum Hall systems \cite{DIVafa-FQHE}. Excitingly, this may lead to tests of complex Chern-Simons theory in the lab.

\end{itemize}

In this short review, we will only be able to touch upon a few of these topics and connections.
We will actually begin in Section \ref{DI.sec:quant} with some basic concepts in complex Chern-Simons theory, including the definition of the Hilbert spaces $\CalH[\Sigma]$ assigned to two-dimensional oriented manifolds. One of the most prominent distinctions between Chern-Simons theory with complex and compact gauge groups is that, in the complex case, these Hilbert spaces are infinite-dimensional. As an illustrative example, we will outline the simple quantization of the torus Hilbert space $\CalH[T^2]$ for gauge group $SL(2,\BC)$, and its dependence on the coupling constants or ``levels'' of the theory. We also review some features of the refined or \emph{equivariant} quantization of Hilbert spaces recently developed by Gukov and Pei \cite{DIGukovPei}.

This prepares us in Section \ref{DI.sec:TQFT} to discuss one of the most fundamental \emph{open} problems in complex Chern-Simons theory: defining the theory as a full TQFT. In essence, this means being able to assign Hilbert spaces $\CalH[\Sigma]$ to any oriented surface and wavefunctions $\CalZ[M]\in \CalH[\pd M]$ to any oriented three-manifold, in such a way that the standard cutting-and-gluing axioms of Atiyah and Segal are obeyed \cite{DIatiyah-1990}. As we shall review, the difficulty with cutting and gluing in complex Chern-Simons theory stems from the infinite-dimensional nature of Hilbert spaces, and the fact that, naively, wavefunctions often vanish or diverge. Some very promising routes to overcoming these difficulties are suggested by embedding complex Chern-Simons theory in string/M-theory, and using additional symmetries to regulate zeroes or infinities  \cite{DICDGS, DIGukovPei, DIPeiKe}. Interestingly, these symmetries are related to categorification of Chern-Simons theory.

In the second half of this review, we then discuss a few relations between complex Chern-Simons theory and the topology and geometry of three-manifolds.
In each case, we view these relations in light of string/M-theory and the 3d-3d correspondence.
In Section \ref{DI.sec:topol}, we will discuss 1) asymptotic expansions of $SL(N,\BC)$ Chern-Simons partition functions, their relation to hyperbolic volumes, and their behavior at large $N$ \emph{vis \`a vis} holography of five-brane systems; 2) state-sum/integral models for $SL(N,\BC)$ Chern-Simons theory, their relation to positive angle structures on ideal triangulations of three-manifolds, and the corresponding implication for positivity of operator dimensions in theories $T[M]$ built from ideal triangulations; 3) the interpretation of the $SL(N,\BC)$ partition function at level $k=0$ as counting surfaces in a three-manifold $M$, BPS operators in $T[M]$, and BPS M2 branes ending on wrapped M5 branes in M-theory (related to recent mathematical work \cite{DIGHHR}). In Section \ref{DI.sec:QM}, we will state some of the observations and conjectures about ``quantum'' modularity in Chern-Simons theory.

We emphasize that the 3d-3d correspondence provides the main link between complex Chern-Simons theory and localization methods in supersymmetric gauge theories, which are the focus of this collection of articles. Particularly relevant are the reviews of T. Dumitrescu (\volcite{DU}) and B. Willett (\volcite{WI}). The basic idea is that whenever $T[M]$ can be explicitly described as (say) a gauge theory, its partition function on spaces such as $S^3$ or $S^3/\BZ_k$ is readily computed by supersymmetric localization. This has led to new formulations and refinements of Chern-Simons partition functions on $M$, which in turn produce invariants of 3-manifolds,
\DIbe
\raisebox{-.4in}{\includegraphics[width=4.5in]{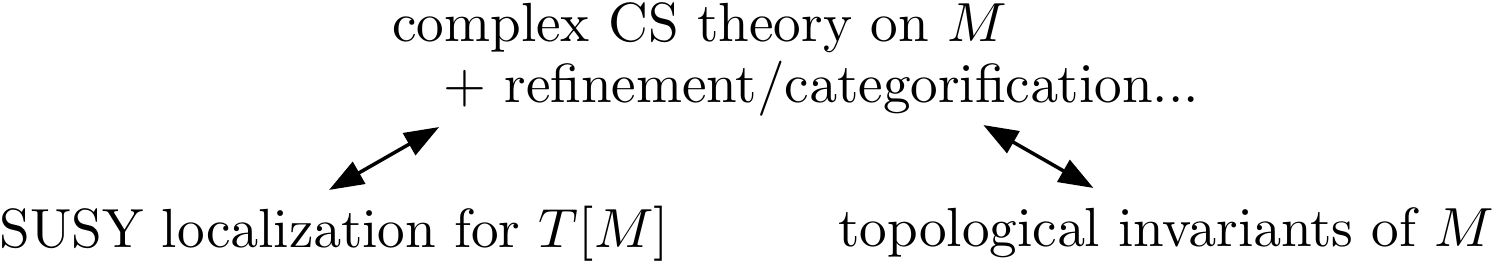}}
\DIee

Unfortunately, we will not say very much about connections of complex Chern-Simons theory to cluster algebras, topological recursion, categorification, gravity, or many other fascinating topics. We hope that some of the references above will guide readers interested in these subjects.

\section{Warmup: quantization of $\CalH(T^2)$}
\label{DI.sec:quant}

To get a feel for the structure of complex Chern-Simons theory, we begin with a (seemingly) elementary exercise: the quantization of the phase space that $SL(2,\BC)$ Chern-Simons theory attaches to a two-torus.

First, some generalities. As discussed in \cite{DIWitten-cx}, the action of complex Chern-Simons theory on a Euclidean three-manifold $M$ takes the form
\DIbe I_{k,s} = \tfrac12(k+is)S_{CS}(\CalA) + \tfrac12(k-is)S_{CS}(\ol\CalA)\,, \label{DI.action} \DIee
where $S_{CS} = \frac{1}{4\pi}\int_M \text{Tr}\big(\CalA\wedge\CalA+\frac23\CalA\wedge\CalA\wedge\CalA\big)$ is the usual Chern-Simons functional. Here $\CalA$ is a connection on an $SL(2,\BC)$ bundle over $M$, and $\ol \CalA$ is its complex conjugate. The group $SL(2,\BC)$ contains $SU(2)$ as its maximal compact subgroup (in fact, as a complex manifold, $SL(2,\BC)\simeq T^*SU(2)$), and on a compact 3-manifold $M$ there can be large gauge transformations $g:M\to SL(2,\BC)$ that wrap nontrivially around the compact $SU(2)$. The path-integral integration measure $\exp(i I_{k,s})$ is invariant under large gauge transformations on a closed $M$ so long as $k\in \BZ$. On the other hand, the coupling constant $s$ is unconstrained.

For a unitary theory --- which in Euclidean space means that partition functions are conjugated under orientation reversal --- the action must be real, which forces $s$ to be real. We will assume this to be true in the quantization below, though eventually in partition functions we will find that we can analytically continue $s$ in a straightforward manner. There also exist exotic unitarity structures with $s$ imaginary \cite{DIWitten-cx}, which will lead to slightly different Hilbert spaces.

As an aside, in the relation to 3d Euclidean gravity with a negative cosmological constant, one identifies the Hermitian and anti-Hermitian parts of $\CalA = w+ie$ as a vielbein $(e)$ and a spin connection $(w)$. The part of the action $I_{k,s}$ proportional to $s$ becomes the usual Einstein-Hilbert action, while the part proportional to $k$ is a gravitational Chern-Simons term \cite{DIWitten-gravCS}. The classical solutions of Chern-Simons theory on a three-maniofld $M$ are flat $SL(2,\BC)$ connections, which become identified with (possibly degenerate) metrics of constant negative curvature, \DIie\ hyperbolic metrics, in 3d gravity.

Geometric quantization of complex Chern-Simons theory on a general surface $\Sigma$ was first discussed in \cite{DIWitten-cx} and recently revisited in \cite{DIAG-HW}, using a holomorphic polarization. A more modern perspective on quantization, based on the topological A-model, appears in \cite{DIDGLZ}, following \cite{DIGW-branes} (see also \cite{DImirrorbranes}). In the case of $\Sigma=T^2$, we can take a more pedestrian approach, following \cite{DID-levelk, DIAK-complexCS}.

The Hilbert space that Chern-Simons theory assigns to any surface $\Sigma$ is a quantization of the classical phase space
\DIbe \CalP[\Sigma] = \CalM_{\rm flat}(SL(2,\BC),\Sigma) \simeq {\rm Hom}(\pi_1(\Sigma),SL(2,\BC))\,. \label{DI.PS} \DIee
This is the space of complex flat connections on $\Sigma$ (modulo gauge transformations), or equivalently, the space of representations of the fundamental group of $\Sigma$ in $SL(2,\BC)$. The space $\CalP[\Sigma]$ is a finite-dimensional complex symplectic variety, possibly singular, equipped with the Atiyah-Bott holomorphic symplectic form
\DIbe \Omega = \int_\Sigma \delta \CalA \wedge \delta \CalA\,. \label{DI.AB} \DIee
Of course, the actual symplectic form we use for quantization should be real; the Chern-Simons action \eqref{DI.action} tells us to take
\DIbe \omega_{k,s} = \frac{k+is}{4\pi}\,\Omega + \frac{k-is}{4\pi}\, \ol\Omega = \frac{k}{2\pi}\text{Re}\,\Omega - \frac{s}{2\pi}\text{Im}\,\Omega\,. \DIee

Famously, the space $\CalP[\Sigma]$ is hyperk\"ahler. It admits an entire $\cp^1$ of complex structures, one of which is singled out in the description \eqref{DI.PS} as a space of complex flat connections. The other complex structures can be made manifest by rewriting $\CalP[\Sigma]$ as Hitchin's moduli space $\CalP[\Sigma]\simeq \CalM_{\rm Hit}(SU(2),\Sigma)$ associated to the compact group $SU(2)$ \cite{DIHitchin-SD}. Similarly, the space $\CalP[\Sigma]$ admits a $\cp^1$ of real symplectic forms, spanned by the hyperk\"ahler triplet $(\omega_I,\omega_J,\omega_K)$, where $\omega_I=\text{Re}\,\Omega$ and $\omega_J = \text{Im}\,\Omega$ above. The third form $\omega_J$ is a K\"ahler form in our chosen complex structure. Notably, $\frac{1}{4\pi^2}\omega_I$ represents a nontrivial \emph{integral} cohomology class in $H^2(\CalP[\Sigma],\BZ)$, while $\omega_J,\omega_K$ are cohomologically trivial. This provides another explanation for the quantization of the level $k$: in geometric quantization, one requires $\omega_{k,s}$ to be the first Chern class of a line bundle, which can only happen if it has integral periods, whence $k\in \BZ$, but $s$ is unconstrained.

Now let us specialize to $\Sigma=T^2$. Flat connections on a torus are determined by the holonomies $\rho_A,\rho_B$ along the $A$ and $B$ cycles, up to $SL(2,\BC)$ conjugation. Since the fundamental group $\pi_1(T^2)=\BZ^2$ is abelian, the holonomies commute and can be simultaneously diagonalized.%
\footnote{More precisely: the holonomies can simultaneously be put in Jordan normal form.} %
Letting $x,y$ denote their eigenvalues, we find
\DIbe \CalP[T^2] = (\BC^*\times \BC^*)/\BZ_2 \simeq \{(x,y)\in (\BC^*)^2\}/\raisebox{-.1cm}{$(x,y)\sim (x^{-1},y^{-1})$}\,.\DIee
The $\BZ_2$ action here is just that of the Weyl group, acting as residual gauge transformations.
The holomorphic symplectic form is
\DIbe \Omega = 2\,\frac{dy}{y}\wedge \frac{dx}{x}\,,\DIee
reflecting the nontrivial intersection of A and B cycles on $T^2$, whence
\DIbe \omega_{k,s} = \frac{k}{\pi}(d\log |y|\wedge d\log |x| - d\arg y\wedge d \arg x) -\frac{s}{\pi}(d\log|y|\wedge d\arg x +d\arg y\wedge d\log |x|)\,. \DIee
As anticipated, the period of $\frac{1}{2\pi}\omega_{k,s}$ on the compact $(S^1\times S^1)/\BZ_2$ cycle in $\CalP[T^2]$ is equal to $k$, and is properly quantized.

In order to diagonalize the real symplectic form $\omega_{k,s}$, we define $b$ to be the complex number with $\text{Re}(b)>0$ and
\DIbe b^2 = \frac{k-is}{k+is}\,, \label{DI.defb}\DIee
and make a change of variables%
\footnote{The new variables absorb some powers of the coupling constants $k,b$, which obfuscates the effect of the classical limit $k\to\infty$, \DIeg\ in \eqref{DI.nmcomm}. It is important to keep in mind that the natural functions on the phase space are still $x,\bar x,y,\bar y$.} %
from $(x,y)\in (\BC^*)^2$ to $(\mu,\nu;m,n)\in \BR^2\times (\BR/2k\BZ)^2$\,:
\DIbe \label{DI.defxy}
 \begin{array}{ll} x = \exp \frac{i\pi }{k}\big(-ib\mu-m\big)\,,\qquad & \bar x =  \exp \frac{i\pi }{k}\big(-ib^{-1}\mu+m\big)\,, \\[.1cm]
y = \exp \frac{i\pi }{k}\big(-ib\nu-n\big)\,,\qquad & \bar y =  \exp \frac{i\pi }{k}\big(-ib^{-1}\nu+n\big)\,.\end{array}
\DIee
Notice that if $s$ is real then $|b|=1$, so $(x,y)$ and $(\bar x,\bar y)$ are complex conjugates, as written. Then the symplectic form collapses to
\DIbe \omega_{k,s} = \frac{\pi}{k} d\nu\wedge d\mu - \frac{\pi}{k}dn\wedge dm\,.\DIee

We proceed to quantize the space as if it were just $\BC^*\times\BC^*$, and restore Weyl invariance later.
The functions $\nu,\mu$ and $n,m$ can simply be quantized to operators with canonical commutation relations
\DIbe -[\bm \nu,\bm \mu] = [\mb n,\mb m] = \frac{k}{i \pi}\,. \label{DI.nmcomm} \DIee
Of course, since $m$ is periodic the spectrum of $\mb n$ is quantized, and vice versa; altogether, both the eigenvalues of both $\mb m$ and $\mb n$ must belong to $\BZ/(2k\BZ)$. The well-defined operators are actually quantizations $\mb x,\bar{\mb x},\mb y,\bar{\mb y}$ of the $\BC^*$-valued functions in \eqref{DI.defxy}, with  $\mb x = \exp \frac{i\pi }{k}\big(-ib\bm\mu-\mb m\big)$, etc. For these we find
\DIbe \mb y \mb x  = q^{\frac12} \mb x\mb y\,,\quad \bar{\mb y}\bar{\mb x}=\tilde q^{\frac12}\bar{\mb x}\bar{\mb y}\,;\qquad \mb x\bar{\mb y} = \bar{\mb y}{\mb x}\,,\quad  \mb y\bar{\mb x} = \bar{\mb x}{\mb y}\,, \label{DI.qtorus} \DIee
with
\DIbe q^{\frac12} := \exp \frac{2\pi i}{k+is} = \exp{\tfrac{i\pi}{k}(b^2+1)}\,,\qquad \tilde q^{\frac12} := \exp \frac{2\pi i}{k-is} = \exp{\tfrac{i\pi}{k}(b^{-2}+1)}\,. \label{DI.qq} \DIee
Thus, abstractly, we see that the quantized operator algebra consists of two independent Weyl algebras (or ``quantum torus'' algebras), one in $\mb x,\mb y$ and one in $\bar{\mb x},\bar{\mb y}$.

There are many equivalent ways to represent the operator algebra on a Hilbert space $\CalH[T^2]$. The simplest is to take
\DIbe \CalH[T^2]_{k,s} = L^2(\BR)\otimes \BC^{2|k|} \simeq \{f(\mu,m)\} \label{DI.Hks} \DIee
to consist of functions of a real variable $\mu$ and an integer $m\in \BZ/(2k\BZ)$. Equivalently, we may take functions of $x$ and $\bar x$. Formally, in geometric quantization, this corresponds to choosing a particular ``real'' polarization --- taking sections of a line bundle $\CalL\to\CalP[\Sigma]$ with $c_1(\CalL)=\omega_{k,s}$ that are covariantly constant with respect to $\nu$ and $n$. The operators $\mb x,\bar{\mb x}$ act on $f(\mu,m)$ as multiplication by $x,\bar x$, while $\mb y,\bar{\mb y}$ are shifts
\DIbe \mb y\,f(\mu,m) = f(\mu+ib,m-1)\,,\qquad \bar{\mb y}\,f(\mu,m) = f(\mu+ib^{-1},m+1)\,.\DIee

To restore Weyl-invariance, we restrict to the $\BZ_2$-invariant part of the Hilbert space, \DIie\ functions that are even
\DIbe f(\mu,m) = f(-\mu,-m)\,. \label{DI.Weyl} \DIee
Correspondingly, we should restrict to a subalgebra of the operator algebra that is invariant under $(\mb x,\bar{\mb x},\mb y,\bar{\mb y})\to(\mb x^{-1},\bar{\mb x}^{-1},\mb y^{-1},\bar{\mb y}^{-1})$. This subalgebra is generated by operators $\mb X=\mb x+\mb x^{-1}$, $\mb Y=\mb y+ \mb y^{-1}$, and $\mb T=\mb{xy} + \mb x^{-1} \mb y^{-1}$ (and their conjugates), which obey
\DIbe \mb X^2 + \mb Y^2 + q^{-\frac12}\mb T^2 = \mb X\mb T\mb Y +2(1+q^{-1})\,. \DIee

When $k=0$, complex Chern-Simons theory still make sense (as long as $s\neq 0$), but the above quantization procedure requires a slight modification \cite{DIDGG-index}. The change of variables \eqref{DI.defxy} does not make sense, and is not necessary, since $\omega_{k=0,s}$ is already diagonalized. Indeed, $\omega_{k=0,s}$ is the canonical symplectic form on $\BC^*\times \BC^*$ when viewed as the cotangent bundle $T^*(S^1\times S^1)$. We expect quantization to produce $\CalH[T^2]_{0,s} = L^2(\BZ\times \BZ)$. To see it, simply write $(x,y)=(e^{\frac{\pi}{s}m+i\theta}, e^{\frac\pi s n+i\phi})$. The symplectic form becomes
\DIbe \omega_{k=0,s} = - dn\wedge d\theta + dm\wedge d\phi\,. \DIee
These canonically-conjugate functions are quantized to operators with $[\bm \theta,\mb n]= - [\bm \phi,\mb m]=i$.
Since $\theta,\phi$ are periodic with period $2\pi$, the eigenvalues of $\mb m,\mb n$ must be integers. We can represent the operator algebra (say) on functions $f(m,n)$ of two integers, such that
\DIbe \label{DI.index-fmn}
 \begin{array}{ll}   \mb x\, f(m,n) = q^{\frac m4} f(m,n-1)\,,\quad &\bar{\mb x}\,f(m,n) = \tilde q^{-\frac m4} f(m,n+1)\,, \\[.1cm]
 \mb y\, f(m,n) = q^{\frac n4} f(m+1,n)\,,\quad &\bar{\mb y}\,f(m,n) = \tilde q^{-\frac n4} f(m-1,n)\,, \end{array}\DIee
where now \eqref{DI.qq} reduces to
\DIbe q^{\frac14} = \tilde q^{-\frac 14} := \exp \frac \pi s\,. \label{DI.qq-index} \DIee
With these new definitions of $q$ and $\tilde q$, the operators satisfy the standard quantum-torus relations \eqref{DI.qtorus}. Alternatively, and equivalently, we may take the Hilbert space to contain functions $g(m,\zeta)$ of an integer $m$ and a phase $\zeta=e^{i\theta}$, with
\DIbe \label{DI.index-fmz}
\begin{array}{ll}   \mb x\, g(m,\zeta) = q^{\frac m4}\zeta\, g(m,\zeta)\,,\quad &\bar{\mb x}\,g(m,\zeta) = \tilde q^{-\frac m4}\zeta^{-1} \,g(m,\zeta)\,. \\[.1cm]
 \mb y\, g(m,\zeta) = q^{\frac n4}\, g(m+1,q^{\frac14}\zeta)\,,\quad &\bar{\mb y}\,g(m,\zeta) = \tilde q^{-\frac n4} \,g(m-1,\tilde q^{-\frac n4} \zeta)\,. \end{array}\DIee
Again, we impose Weyl-invariance at the end by restricting to even functions $f(m,n)=f(-m,-n)$ or $g(m,\zeta)=g(-m,\zeta^{-1})$.

\subsection{Equivariant quantization}
\label{DI.sec:equiv}

There are several things to notice about \eqref{DI.Hks}. Perhaps the most salient is that, unlike in the case of Chern-Simons theory with compact gauge group \cite{DIWit-Jones, DIaxelrod-witten}, the Hilbert space is infinite-dimensional. This is no surprise, since it comes from quantizing a noncompact phase space. One may also recognize the $\BC^{2k}$ factor as being related to the standard Hilbert space for $SU(2)$ theory at (bare) level $k$. Indeed, if we ignore $\mu$ and consider functions $f(m)$ of an integer $m\in \BZ/(2k\BZ)$, such that $f(m)=f(-m)$ as in \eqref{DI.Weyl}, we find exactly $|k|+1$ independent values $f(0),f(1),...,f(|k|)$ that determine a state in the finite-dimensional Hilbert space of $SU(2)$ Chern-Simons.

What is less obvious, in particular for general $\Sigma$, is that the infinite-dimensional Hilbert space of complex Chern-Simons theory admits an additional $U(1)_t$ symmetry, introduced in \cite{DIGukovPei}. The graded components of $\CalH[\Sigma]$ (\DIie\ the subspaces of fixed $U(1)_t$ charge) turn out to be finite-dimensional, and in particular the subspace of zero charge is just the familiar $SU(2)$ Hilbert space.

The extra $U(1)_t$ symmetry comes from viewing $\CalP[\CalM]\simeq \CalM_{\rm Hit}(SU(2),\Sigma)$ as the Hitchin moduli space. There is a canonical $U(1)_t$ metric isometry of the Hitchin moduli space that rotates the $\cp^1$ of complex structures about an axis. In particular, it rotates $\omega_J=\text{Im}(\Omega)$ and $\omega_K$ into each other. This is an isometry of our quantization problem at least when $s=0$, since it preserves $\omega_{k,s=0}$, and leads to the desired symmetry of $\CalH[\Sigma]_{k,0}$.
(Since the Hilbert space, abstractly, does not depend on $s$, one might then hope to endow even spaces at $s\neq 0$ with the symmetry.)

In the case $\Sigma=T^2$, it is easy to describe the $U(1)_t$ symmetry: when we view $\CalP[T^2]\simeq T^*(S^1\times S^1)/\BZ_2\approx T^*\CalM_{\rm flat}(SU(2),T^2)$ as the cotangent bundle of the space of flat $SU(2)$ connections, $U(1)_t$ simply rotates the cotangent fibers. The $U(1)_t$-invariant subspace of $\CalH[T^2]_{k,s=0}$ simply consists of the functions $f(\mu,m)$ that are independent of $\mu$ --- \DIie\ the $SU(2)$ Hilbert space we found above. The full $U(1)_t$ action is trickier to describe in the polarization we are using. Roughly, one observes that at $s=0$ and $b=1$ the variables $x,\bar x$ can be written as
\DIbe x = z \eta^{-m}\,,\quad \bar x = z\eta^{m}\,, \DIee
where $z = \exp(\pi \mu/k)$ and $\eta=\exp(i\pi/k)$. Then, on functions of $f(z,m)$ that are analytic in $z$, the $U(1)_t$ symmetry just acts as rotations $z\to e^{i\theta}z$. The subspaces of fixed $U(1)_t$ weight contain monomials in $z$. After imposing Weyl invariance, the graded dimension of the Hilbert space becomes
\DIbe \begin{array}{ll} \text{dim}_{U(1)_t} \CalH[T^2]_{k,s=0} &:= \sum_{w\in \BZ} t^w \text{dim}\CalH[T^2]_{k,s=0}^{\text{weight $w$}} = |k|+1 + 2|k|(t+t^2+t^3+...) \\[.1cm]
&= \ds |k|+1 + \frac{2|k|t}{1-t}\,. \end{array} \label{DI.grdim} \DIee

\subsection{Holomorphic polarizations and CFT}

Often in geometric quantization of Chern-Simons theory, one uses a holomorphic polarization instead of the real polarization above. In Chern-Simons theory with compact gauge group (\DIcf\ \cite{DIaxelrod-witten}), this means to choose a complex structure `$\tau$' on a surface $\Sigma$, to write the connection one-form as $A = A_zdz+A_{\bar z}d\bar z$ in local complex coordinates, and, when quantizing, to define the Hilbert space to consist of sections of the line bundle $\CalL\to \CalP[\Sigma]$ that are covariantly constant with respect to $A_{\bar z}$.

Naively, it may appear that choosing such a complex polarization needlessly complicates the problem. However, a complex polarization has three great advantages. First, it allows one to ask analyze the Hilbert space varies with the choice of complex structure. Locally the variation is trivial, expressed formally by saying that the bundle of Hilbert spaces over the space of complex structures (\DIie\ over the Teichm\"uller space of $\Sigma$) has a projectively flat connection. However, globally, one derives an action of the mapping class group of $\Sigma$ on the Chern-Simons Hilbert space. Second, and related to this idea, a holomorphic polarization allows one to identify the Hilbert space of Chern-Simons theory with the space of conformal blocks in a particular boundary CFT. In the case of compact Chern-Simons theory, the boundary CFT is a famously WZW model \cite{DIWit-Jones}. The projectively flat connection on Teichm\"uller space is the Knizhnik-Zamolodchikov connection of the CFT. Finally, for generic surfaces $\Sigma$ a real polarization as above simply isn't available! Thus, using a holomorphic polarization is the only way to go.

In the case of complex Chern-Simons theory, there are actually multiple choices of complex polarizations. If we write a complex connection and its conjugate as $\CalA = \CalA_z dz+\CalA_{\bar z}d\bar z$, $\ol \CalA = \ol\CalA_z dz+\ol\CalA_{\bar z}d\bar z$, then we can ask that sections of $\CalL$ be covariantly constant with respect to
\DIbe\text{(A)} \quad  \CalA_{\bar z}\,,\quad \ol \CalA_{z} \qquad \text{or}\qquad \text{(B)} \;\; \text{(some components of)} \; \;  \ol\CalA_{\bar z}\,,\quad \ol \CalA_{z}\,.
\DIee
The first choice was analyzed by Witten \cite{DIWitten-cx}, and leads to a boundary CFT whose conformal blocks must contain contributions from both chiral and anti-chiral sectors. This polarization plays a central role in the relation between $SL(2,\BC)$ Chern-Simons theory and quantum Hall systems \cite{DIVafa-FQHE, DIDGV-FQHE}. The second choice is related to Liouville theory coupled to parafermions \cite{DINT-para, DICJ-AGT}.

\section{Complex Chern-Simons theory as a TQFT?}
\label{DI.sec:TQFT}

Now, having seen very explicitly that complex Chern-Simons Hilbert spaces are infinite-dimensional (and exactly how they're infinite-dimensional), let us think a bit about the properties of partition functions.

For a closed three-manifold $M$, it is expected that the complex Chern-Simons partition function takes the form
\DIbe \CalZ[M]_{k,s} = \sum_{\text{flat $\alpha$}} \frac{1}{|\text{Stab}(\alpha)|} \CalB_\alpha(q^{\frac12})\widetilde\CalB_\alpha(\tilde q^{\frac12}) \,. \label{DI.fact} \DIee
The sum here is over flat complex connections $\alpha$ on $M$, which are the critical points of the Chern-Simons path integral; and $\CalB_\alpha$, $\widetilde\CalB_\alpha$ are holomorphic and antiholomorphic contributions to the path integral from quantum fluctuations around the critical point, with $q,\tilde q$ defined by \eqref{DI.qq}. The prefactor $|\text{Stab}(\alpha)|$ is the volume of the stabilizer of $\alpha$, \DIie\ the volume of the subgroup of the gauge group that preserves a particular flat connection.

One certainly expects such a formula to be valid perturbatively, due to standard properties of path integrals in quantum field theory. (The perturbative version of \eqref{DI.fact} formed the basis for the physical explanation of the Volume Conjecture in \cite{DIgukov-2003}.) It was argued in \cite{DIWit-anal}, however, that the formula is actually valid non-perturbatively as well. Roughly, one should think of \eqref{DI.fact} as expanding the integration cycle in the Chern-Simons path integral into a sum of integration cycles $\Gamma_\alpha$ defined by gradient flow off of each critical point with respect to the Chern-Simons action. The $\Gamma_\alpha$ can further be written as products%
\footnote{In general, there could be a nontrivial matrix $n_{\alpha\beta}$ connecting holomorphic and anti-holomorphic sectors. However, when one considers unitary Chern-Simons theory with an integration contour $\Gamma$ along which $\ol\CalA$ is honestly the conjugate of $\CalA$, the matrix is just the identity.} %
of cycles $\gamma_\alpha\times \widetilde\gamma_\alpha$ in the space of (holomorphic)$\times$(anti-holomorphic) connections, leading to the factorization $\CalB_\alpha(q^{\frac12})\widetilde\CalB_\alpha(\tilde q^{\frac12})$.\,%
\footnote{The 3d-3d correspondence relates holomorphic-antiholomorphic factorization in complex Chern-Simons theory to a rather nontrivial statement about lens-space partition functions of 3d $\CalN=2$ theories $T[M]$~\cite{DIBDP-blocks}.}

For a three-manifold with boundary $\Sigma$, the same type of formula holds after properly accounting for boundary conditions. In particular, the sum is over flat connections with a fixed behavior at $\Sigma$, and each summand becomes a wavefunction  in $\CalH[\Sigma]_{k,s}$. For example if $\Sigma=T^2$ is a torus, we would fix A-cycle holonomy eigenvalues $x,\bar x$ of a flat connection as in Section \ref{DI.sec:quant}, and the individual wavefunctions would have the factorized form $\CalB_\alpha(x,q^{\frac12})\widetilde\CalB_\alpha(\bar x,\tilde q^{\frac12})$.

What can we learn from \eqref{DI.fact}? In the best-case scenario, there is a finite number of flat connections on $M$, and all the flat connections have finite-volume stabilizers. This would lead to a finite, well-defined (in principle) $\CalZ[M]_{k,s}$. In contrast:
\begin{itemize}
\item If a flat connection $\alpha$ is isolated but its stabilizer has infinite volume, its contribution to \eqref{DI.fact} vanishes.
\item If flat connections come in a continuous family on which the Chern-Simons action is constant, then the contribution to \eqref{DI.fact} can be infinite.
\end{itemize}
Unfortunately, the best-case scenario never holds, and both of these potentially bad situations can arise. We consider some examples.

If $M$ is hyperbolic (meaning that it admits a hyperbolic metric), it is expected that there are a finite number of $SL(2,\BC)$ flat connections on $M$.
Intuitively, for hyperbolic $M$ the fundamental group $\pi_1(M)$ is sufficiently complicated that the representations $\pi_1(M)\to SL(2,\BC)$ are isolated.%
\footnote{On hyperbolic $M$, one particular flat $SL(2,\BC)$ connection --- the one corresponding to the global hyperbolic metric --- is well known to be isolated \cite{DImostow-1973, DIthurston-1980}. Computational experiments suggest that in fact all flat connections $SL(2,\BC)$ are isolated, but there exists no general proof of this statement.} %
Then the partition function is finite. As we discuss in Section \ref{DI.sec:angle}, the partition function can actually be computed.
However, there is always at least one flat connection $\alpha_{\rm abel}$ whose holonomies belong to the maximal torus $GL(1,\BC)\subset SL(2,\BC)$. The stabilizer of $\alpha_{\rm abel}$ contains constant $GL(1,\BC)$-valued gauge transformations; since $GL(1,\BC)$ has infinite volume, $\alpha_{\rm abel}$ does not contribute at all to the partition function. This becomes hugely problematic when trying to formulate complex Chern-Simons as a TQFT, as all flat connections must be accounted for during cutting and gluing \cite{DICDGS}.

There are some simple manifolds whose fundamental group $\pi_1(M)$ is abelian. For example, if $M$ is a lens space $L(p,r)\simeq S^3/\BZ_p$, the fundamental group is $\BZ_p$. In this case, every single flat connection has holonomy in the maximal torus of the gauge group, the volume of the stabilizer is always infinite, and $\CalZ[M]$ vanishes identically.

In the opposite extreme are manifolds $M$ on which the flat connections are not isolated. In this case, we expect that $\CalZ[M]$ diverges.
For example, consider $M=\Sigma\times S^1$. Then  $\CalM_{\rm flat}(G_\BC,M) = \CalM_{\rm flat}(G_\BC,\Sigma)\times T_\BC$ (where $T_\BC$ is the maximal torus of $G_\BC$). The partition function is
\DIbe \CalZ[M]_{k,s} = \text{Tr}_{\CalH[M]_{k,s}} 1\hspace{-.15cm}1 = \text{dim}\,\CalH[M]_{k,s} = \infty\,. \label{DI.tr} \DIee

Both the zeroes and infinities appearing here obstruct the definition of consistent cutting and gluing rules needed to make Chern-Simons theory a TQFT. The zeroes and infinities have to be regularized. While no systematic approach to regularization has been formulated so far, there are exist several promising and exciting proposals. Almost all of them are motivated by string/M-theory and the 3d-3d correspondence.

\subsection{Symmetries, regularizations, and the 3d-3d correspondence}
\label{DI.sec:sym}

In principle, the 3d-3d correspondence itself may suffice to resolve the difficulties with cutting and gluing in complex Chern-Simons theory.%
\footnote{We will not review the 3d-3d correspondence here. For recent reviews and discussions, especially in the context of Chern-Simons theory and TQFT, see \cite{DID-volume, DICDGS, DIGukovPei, DIPeiKe} as well as the related \cite{DIYamazaki-defects}.} %
The correspondence assigns to a closed 3-manifold $M$ and a Lie algebra $\mathfrak g$ of type $ADE$ (and a bit of extra discrete data) a three-dimensional field theory $T_G[M]$ with the property that that
\DIbe \begin{array}{l} \text{partition function of $T_G[M]$ on squashed lens space $L(k,1)_b$} \\[.1cm]
\qquad =\text{$G_\BC$ Chern-Simons partition function $\CalZ[M]_{k,s}$ with $is=k\frac{1-b^2}{1+b^2}$.}
\end{array}\DIee
where $\mathfrak g=\text{Lie}(G)$.
When $G_\BC=SL(N,\BC)$, the theory $T_{G_\BC}[M]$ is the effective low-energy worldvolume theory of $N$ M5 branes compactified on $M$; the branes wrap $M\times \BR^3$ in the M-theory geometry $T^*M\times \BR^5$.
Similarly, the correspondence assigns to a three-manifold $M$ with boundary $\Sigma$ a boundary condition $T_G[M]$ for the four-dimensional theory $T_G[\Sigma]$ of class $\CalS$ \cite{DIGaiotto-dualities, DIGMN}.

Typically $T_G[M]$ is an $\CalN=2$ superconformal theory, though both supersymmetry and conformal invariance might be broken. Unfortunately, it is not completely understood what conditions on $M$ guarantee $\CalN=2$ superconformal theories. We assume for the present heuristic argument that we do have $\CalN=2$ superconformal theories.

The basic idea, then, would be to replace $G_\BC$ Chern-Simons theory on $M$ with $T_{G_\BC}[M]$, which is a much more powerful object. Even when the Chern-Simons partition functions $\CalZ[M]_{k,s}$ (equivalently, lens-space partition functions of $T_G[M]$) are ill-defined, the theory $T_G[M]$ itself should still make sense. Moreover, the theories $T_G[M]$ obey cutting and gluing rules. Gluing $M=N_1\cup_\Sigma N_2$ corresponds to ``sandwiching'' the four-dimensional theory $T_G[\Sigma]$ between boundary conditions $T_G[N_1]$ and $T_G[N_2]$, and colliding  the boundaries together to produce a new effective theory $T_G[N_1\cup_\Sigma N_2]$. In this way, we reproduce the structure of a three-dimensional TQFT. If we should ever want to recover Chern-Simons partition functions, we just place the superconformal theories on a lens space $L(k,1)_b$.

There are two practical difficulties with this proposal that will hopefully be overcome soon. First, the full theories $T_G[M]$ are not actually known for most manifolds, for any nonabelian $G$. A construction using ideal triangulations was outlined in \cite{DIDGG, DIDGG-Kdec} (also \cite{DICCV, DICordova-tangles}); however, that construction produces subsectors of the full theories $T_G[M]$ that are missing some branches of vacua, the same way partition functions of Chern-Simons theory on hyperbolic manifolds are ``missing'' abelian flat connections. Examples of complete theories $T_G[M]$ for a handful of manifolds (including a hyperbolic one) were postulated in \cite{DICDGS}, and theories for lens spaces $M=L(p,r)$ were studied in \cite{DICDGS, DIGukovPei, DIPeiKe}.

The second difficulty, or potential shortcoming, is that zeroes and infinities still remain in the actual Chern-Simons partition functions. Here, however, another solution presents itself: the theories $T_G[M]$ often have extra symmetries; and parameters associated to these symmetries (twisted masses or fugacities) can be used to refine the squashed-lens-space $L(k,1)_b$ partition functions of $T_G[M]$ --- thus literally regularizing the Chern-Simons zeroes and infinities.

We already met one such symmetry in Section \ref{DI.sec:equiv}: the $U(1)_t$ that gave an equivariant quantization of the Hilbert space $\CalH[\Sigma]_{k,s=0}$. Via the 3d-3d correspondence, this Hilbert space is mapped to the BPS Hilbert space of the 4d class-$\CalS$ theory $T_G[\Sigma]$ on $\BR\times L(k,1)_{b=1}$. This 4d theory has an additional R-symmetry $U(1)_t$ that commutes with the supercharge used to define the ``BPS'' Hilbert space, and provides the $U(1)_t$ grading.

Similarly, the three-dimensional theory $T_G[\Sigma\times S^1]$ (obtained by compactifying $T_G[\Sigma]$ on a circle) has $\CalN=4$ rather than $\CalN=2$ supersymmetry. The larger R-symmetry group of the $\CalN=4$ theory contains $U(1)_t$, and including its twisted mass in partition functions leads to finite answers that encapsulate the graded dimension \eqref{DI.grdim}.

Theories $T_G[M]$ for Seifert-fibered three-manifolds $M$ should also retain this $U(1)_t$ symmetry. In this case, it can ultimately be traced back to an exceptional isometry%
\footnote{This very same symmetry played a central role in defining refined (compact) Chern-Simons theory on Seifert-fibered manifolds \cite{DIAS-refinedCS}.} %
 of the $M$-theory geometry $T^*M\times \BR^5$. An simple example of such a manifold is a lens space $M=L(p,r)$, whose refined partition functions were analyzed in \cite{DIPeiKe}, and put precisely into the factorized form \eqref{DI.fact} --- with the factors $1/|\text{Stab}(\alpha)| = 1/\infty$ now regularized.

When $M$ is generic (\DIeg\ hyperbolic) this exceptional $U(1)_t$ symmetry is, unfortunately, absent. It was nevertheless proposed in \cite{DICDGS} that there exists yet \emph{another} symmetry $U(1)_{t'}$ in any theory $T_G[M]$, related to the standard $U(1)$ R-symmetry of three-dimensional $\CalN=2$ theories --- as well as to categorification of colored knot polynomials. This $U(1)_{t'}$ was used to regularize Chern-Simons partition functions for the trefoil and figure-eight knot complements (Seifert-fibered and hyperbolic manifolds, respectively), producing sums of the form \eqref{DI.fact} that included all flat connections, even abelian ones.

\section{Three connections to three-manifold topology}
\label{DI.sec:topol}

As discussed in Section \ref{DI.sec:TQFT}, the partition function of Chern-Simons theory with gauge group $G_\BC$ will take a finite, well-defined value on manifolds $M$ that only admit finitely many flat $G_\BC$ connections.  For $G_\BC=SL(2,\BC)$ and possibly $G_\BC=SL(N,\BC)$, hyperbolic manifolds are expected to be of this type. (Indeed, there exist systematic computations of partition functions for hyperbolic manifolds with boundary of genus $\geq 1$.) One may then try to relate properties of the partition function with the topology of $M$. We proceed to outline some of the more striking relations. In each case, M-theory and/or the 3d-3d correspondence provides valuable insight.

\subsection{Hyperbolic volumes, twisted torsion, and large $N$}
\label{DI.sec:vol}

The most fundamental relation between complex Chern-Simons theory and hyperbolic geometry has been understood for a long time, and concerns the semi-classical asymptotic expansion of partition functions \cite{DIWitten-gravCS} (see also \cite{DIgukov-2003, DIGukovMurakami}). It is easiest to formulate it first in terms of the ``holomorphic blocks'' $\CalB_\alpha,\,\widetilde\CalB_\alpha$ appearing in \eqref{DI.fact}, labelled by flat connections $\alpha$\,:
\DIbe \label{DI.B-asymp}
\begin{array}{ll} \text{as $k+is\to\infty$ or $q^{\frac12} = e^{\frac{2\pi i}{k+is}}\to 1$}\,,\qquad & \ds \CalB_\alpha(q^{\frac12}) \sim \sqrt{\frac{4\pi^3}{\tau(\alpha)}}e^{\frac{i(k+is)}{8\pi}S_{CS}(\alpha)}\,; \\[.2cm]
\text{as $k-is\to\infty$ or $\tilde q^{\frac12} = e^{\frac{2\pi i}{k-is}}\to 1$}\,,\qquad & \ds \widetilde \CalB_\alpha(\tilde q^{\frac12}) \sim \sqrt{\frac{4\pi^3}{\tau(\alpha)^*}}e^{\frac{i(k-is)}{8\pi}S_{CS}(\alpha)^*}\,;
\end{array}
\DIee
where $S_{CS}(\alpha)$ is the classical Chern-Simons action evaluated on a particular flat connection and $\tau(\alpha)$ is the analytic Ray-Singer torsion twisted by the flat connection $\alpha$ \cite{DIRaySinger}. This is the standard result expected from Chern-Simons perturbation theory \cite{DIBW-cx}. In the presence of a boundary, the classical action and torsion on the RHS depend on the choice of boundary conditions (boundary holonomies) for $\alpha$. From \eqref{DI.fact}, it then follows that if both $k+is\to\infty$ and $k-is\to \infty$
\DIbe \CalZ[M]_{k,s} \sim \sum_{\text{flat $\alpha$}} \frac{1}{|\text{Stab}(\alpha)|} \frac{4\pi^3}{|\tau(\alpha)|} e^{\frac{ik}{4\pi}\text{Re}\,S_{CS}(\alpha)-\frac{is}{4\pi}\text{Im}\,S_{CS}(\alpha)}\,. \label{DI.sum-asymp} \DIee
The formula is particularly meaningful if $s$ is analytically continued to (say) positive imaginary values. Then each term in \eqref{DI.sum-asymp} shows exponential growth or decay at large $|s|$, controlled by $\text{Im}\,S_{CS}(\alpha)$.

Now suppose that $M$ is hyperbolic, meaning that it admits a hyperbolic metric. The hyperbolic metric is unique (given suitable boundary conditions) and is a topological invariant of $M$ \cite{DImostow-1973, DIthurston-1980}. Moreover, the vielbein and spin connection of the hyperbolic metric can be rewritten as a flat $SL(2,\BC)$ connection $\alpha_{\rm hyp} = w+ie$, with the property that $\text{Im}\,S_{CS}(\alpha_{\rm hyp}) = \text{Vol}(M)$ is the hyperbolic volume of $M$. The real part $\text{Re}\,S_{CS}(\alpha_{\rm hyp})$ is known as the Chern-Simons invariant of the hyperbolic structure, and provides a natural complexification of the hyperbolic volume \cite{DICheegerSimons, DIChernSimons}. It is also useful to note that the connection $\alpha_{\rm hyp}$ necessarily has a trivial stabilizer -- the connection is fundamentally non-abelian.

Therefore, if the gauge group is $SL(2,\BC)$ and $M$ is hyperbolic, the sum \eqref{DI.sum-asymp} contains a term that is controlled by the hyperbolic volume of $M$. Typically, $\text{Im}\,S_{CS}(\alpha_{\rm hyp})$ is larger than the ``volume'' of any other flat connection, and the entire sum \eqref{DI.sum-asymp} is dominated by the hyperbolic volume. (It is expected that $\text{Im}\,S_{CS}(\alpha_{\rm hyp})$ always dominates, but no general result of this type has been proven.)

It is often useful to strip off the holomorphic part of the asymptotic expansion. This can be done by taking a \emph{singular} limit: we fix $k=1$, analytically continue $s$ to imaginary values, and send $s\to - ik=-i$. This has the effect of sending $q\to 1$ but $\tilde q\to 0$, which trivializes the anti-holomorphic blocks, $\widetilde \CalB_\alpha(\tilde q^{\frac12})\to 1$. In this singular limit, we expect
\DIbe \CalZ[M]_{1,s} \sim \sum_{\text{flat $\alpha$}} \frac{1}{|\text{Stab}(\alpha)|} \sqrt{\frac{4\pi^3}{\tau(\alpha)}} e^{-\frac{1}{\hbar}S_{CS}(\alpha)}\qquad\text{as}\quad \hbar=2\pi i\frac{1-is}{1+is}\to 0\,. \label{DI.sing1} \DIee
This sort of limit played a major role in early analyses of partition functions for complex Chern-Simons theory \cite{DIhikami-2006, DIDGLZ}, though it was not realized at the time that the partition functions (derived from quantum Teichm\"uller theory) being analyzed had fixed level $k=1$.

The fact that the perturbative expansion of complex Chern-Simons theory (or individual holomorphic blocks, as in \eqref{DI.B-asymp}) contains geometric invariants of $M$ played a major role in providing a physical justification for the Volume Conjecture, and generalizing it. The original Volume Conjecture \cite{DIkashaev-1997, DIMur-Mur} claims that a particular double-scaling limit of colored Jones polynomials of a knot $K\subset S^3$ leads to exponential growth, controlled by the hyperbolic volume of the knot complement $\text{Vol}(S^3\bs K)$. This was justified in \cite{DIgukov-2003} by embedding $SU(2)$ Chern-Simons theory (which computes colored Jones polynomials) into a holomorphic sector of $SL(2,\BC)$ Chern-Simons theory, and arguing that the asymptotic expansions of $SU(2)$ and (holomorphic) $SL(2,\BC)$ theories should coincide. The argument immediately led to generalizations, involving higher-order terms in the asymptotic expansion and a dependence on boundary conditions, which have been carefully checked in many computations, \DIcf\ \cite{DIGukovMurakami, DIDGLZ, DIDG-quantumNZ}.

It is also interesting to consider ``large-$N$'' limits in complex Chern-Simons theory, taking the gauge group to be $SL(N,\BC)$ and sending $N\to\infty$. Physically, such limits are most conveniently studied by realizing Chern-Simons theory on a stack of $N$ M5 branes (Section \ref{DI.sec:sym}), and using AdS/CFT or large-$N$ duality. A study of the five-brane system \cite{DIHenningsonSkenderis, DIHMM-anomalies, DIGKW-5branes} predicts that the leading asymptotic growth of the partition function $\CalZ[M]_{k,s}$ as in \eqref{DI.sum-asymp} is scales as $N^3\text{Vol}(M)$ at large $N$ \cite{DIDGG-Kdec}. This is not surprising: the hyperbolic flat $SL(2,\BC)$ connection can be embedded into $SL(N,\BC)$ by using the $N$-dimensional representation $\rho_N:SL(2,\BC)\to SL(N,\BC)$, and the Chern-Simons functional evaluates to $\sim N^3\text{Vol}(M)$ on $\rho_N(\alpha_{\rm hyp})$, \DIcf\ \cite{DIGTZ-slN}. As long as \eqref{DI.sum-asymp} is dominated by the flat connection $\rho_N(\alpha_{\rm hyp})$ for any $N$, one quickly recovers the scaling prediction.
 Much more non-trivially, the M-theory analysis predicts that the logarithm of the \emph{torsion} $\tau(\rho_N(\alpha_{\rm hyp}))$ will grow as $N^3\text{Vol}(M)$ \cite{DIMPS-S3b, DIGKL-largeN} at large $N$ as well. This latter result was recently proved by Porti and Menal-Ferrer \cite{DIPorti-largeN}.

\subsection{State-integral models and angle structures}
\label{DI.sec:angle}

When $M= S^3\bs K$ is an oriented hyperbolic knot or link complement (or, more generally, an oriented hyperbolic manifold with non-empty boundary of genus $\geq 1$), there exists a systematic construction of $SL(2,\BC)$ Chern-Simons partition functions $\CalZ[M]_{k,s}$ for all levels $k,s$. The full definition of these partition functions appears in \cite{DIGar-index, DIGHRS-index} for $k=0$, and \cite{DID-levelk, DIAK-complexCS} for $k\neq 0$, respectively. It is a culmination of much previous work, including  \cite{DIhikami-2006,DIDGLZ,DIDimofte-QRS, DIKashAnd, DIAK-new} for $k=1$; \cite{DIDGG-index}  for $k=0$ (and indirectly \cite{DIKim-index, DIIY-index, DIKW-index}); and (indirectly, via 3d-3d correspondence) \cite{DIBNY-lens, DIIMY-fact} for $k> 1$. The definition extends to $SL(N,\BC)$ using techniques of~\cite{DIDGG-Kdec}.

\begin{figure}[htb]
\centering
\includegraphics[width=5in]{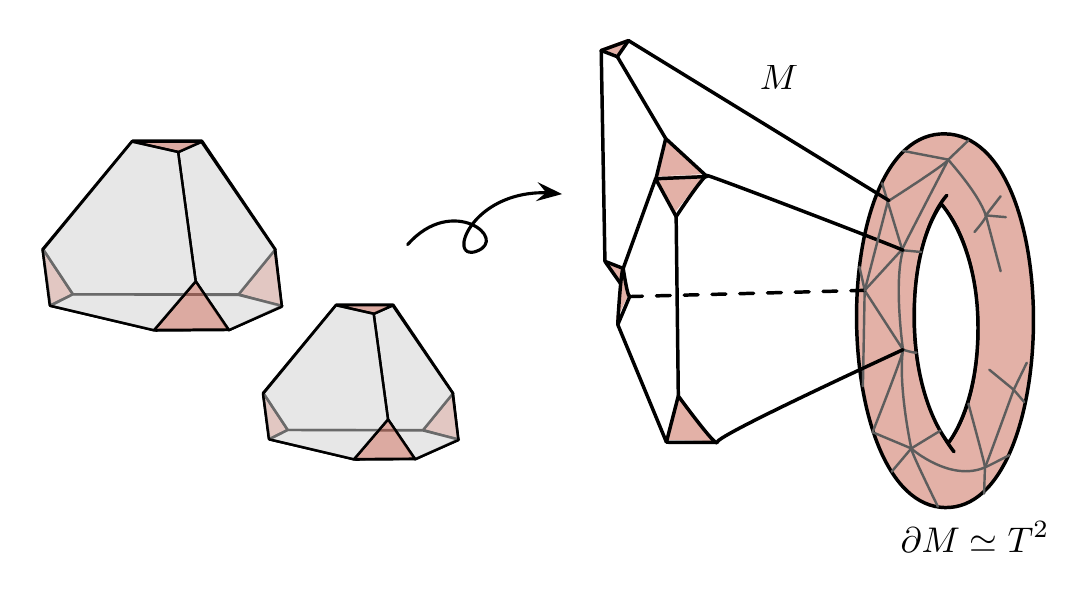}
\caption{Ideal (\DIie\ truncated) tetrahedra, glued together along their large hexagonal faces to form an ideal triangulation of a knot complement $M$. The small triangular faces at truncated vertices become part of the boundary of $M$.}
\label{DI.fig:triang}
\end{figure}

The construction of these partition functions uses a topological ideal triangulation of the three-manifold $M=\cup_{i=1}^N\Delta_i$. This is a tiling of $M$ by truncated tetrahedra $\Delta_i$, as in Figure~\ref{DI.fig:triang}, such that 1) various pairs of large hexagonal faces are glued together; but 2) the small triangles at the truncated vertices are left untouched, and become part of the boundary~$\pd M$.

One then proceeds in standard TQFT fashion. For gauge group $SL(2,\BC)$ (or more precisely, $PSL(2,\BC)$), the boundary of each tetrahedron is assigned a Hilbert space $\CalH[\pd\Delta]_{k,s}\simeq L^2(\BR)\otimes \BC^k$. The Hilbert space comes from quantizing an open subset of the space of flat connections on $\pd\Delta$, viewed as a four-punctured sphere; in this case $\CalP[\pd\Delta]^{\rm open}\simeq \BC^*\times \BC^*$, and the quantization of Section \ref{DI.sec:quant} applies in a straightforward manner, with no Weyl quotient. Each tetrahedron is assigned a canonical partition function $\CalZ[\Delta]_{k,s}\in \CalH[\pd\Delta]_{k,s}$, which has the form of a ``quantum dilogarithm'' function; for $|k|\geq 1$, this is
\DIbe \CalZ[\Delta_i]_{k,s}(\mu_i,m_i) = \prod_{r=0}^\infty \frac{1-q^{1+r}z_i^{-1}}{1-\tilde q^{-r} \bar z_i^{-1}}\,, \label{DI.qdl} \DIee
with $q,\tilde q$ as in \eqref{DI.qq} or \eqref{DI.qq-index} and $z_i=e^{\frac{2\pi b}{k}\mu_i-\frac{2\pi i}{k}m_i},\, \bar z_i=e^{\frac{2\pi b^{-1}}{k}\mu_i+\frac{2\pi i}{k}m_i}$ as in \eqref{DI.defxy}. (For $k=0$, see Section \ref{DI.sec:M2}.)

The partition function $\CalZ[M]_{k,s}$ is then obtained by taking a product $\prod_{i=1}^N \CalZ[\Delta_i]_{k,s}(\mu_i,m_i)$, and integrating out all pairs of variables $(\mu_i,m_i)$ associated to the interior of $M$, leaving behind some variables associated to the boundary. It is slightly tricky to describe this operation in precise terms, because the tetrahedron Hilbert space $\CalH[\pd\Delta]_{k,s}$ does not easily factorize into contributions from the tetrahedron's four large faces.%
\footnote{The Hilbert space \emph{can} be factorized by introducing some redundant degrees of freedom \cite{DIKashAnd}, in a manner directly analogous to Kashaev's quantization of Teichm\"uller space \cite{DIKash-Teich}.} %
The right prescription comes from viewing gluing in TQFT somewhat more globally, as a quantum symplectic reduction. It turns out that the combinatorics of ideal triangulations have some fundamental symplectic properties, first discovered by Neumann and Zagier \cite{DINZ, DINeumann-combinatorics}, that allow the quantum symplectic reduction to be defined.

As an example, take $M=S^3\bs K$ to be a knot complement. In this case, there is a canonical ``A-cycle''  on the boundary torus $\pd M=T^2$, defined by a small loop linking the knot $K$ in $S^3$ (called the meridian cycle); and there is a canonical boundary condition for $SL(2,\BC)$ connections, defined by requiring their holonomy along the meridian cycle to have trivial eigenvalues. (Crucially, this does not require the actual holonomy to be trivial; in $SL(2,\BC)$ there are parabolic matrices $\left(\begin{smallmatrix}1&1\\0&1\end{smallmatrix}\right)$ with trivial eigenvalues.)
In the notation of Section \ref{DI.sec:quant}, the boundary condition sets $x=\bar x=1$, or restricts functions $f(\mu,m)\in \CalH[T^2]_{k,s}$ to their values at $\mu=m=0$. Suppose that the knot complement is glued from $N$ tetrahedra $\Delta_i$. The combinatorics of the triangulation define a ``Neumann-Zagier datum''
\DIbe \mb A\,,\; \mb B \;\in \text{Mat}_{N\times N}(\BZ)\,,\qquad \nu\,, \in \BZ^{N} \label{DI.NZ}
\DIee
consisting of two $N\times N$ integer matrices $\mb A,\,\mb B$ (that encode adjacency relations for edges of the tetrahedra and satisfy the symplectic property $\mb A\mb B^T-\mb B\mb A^T=I_{N\times N}$) and a vector $\nu$ of $N$ integers.
 A precise definition is given in \cite{DIDG-quantumNZ}. Then the partition function $\CalZ[M]_{k,s}=\CalZ[M]_{k,s}(0,0)$ with the canonical boundary condition takes the concise form \cite{DIDG-levelk}
\DIbe \CalZ[M]_{k,s} =   \frac{C}{k^N\sqrt{\text{det}{\mb B}}} \sum_{m\in (\BZ/k\BZ)^N} \int d^N\!\mu\,
(-\zeta^{\frac12})^{m\mb B^{-1}\mb A m}e^{-\frac{i\pi}{k}\mu\mb B^{-1}\mb A\mu -\frac{\pi}{k}(b+b^{-1})\mu\mb B^{-1}\nu} \prod_{i=1}^N \CalZ[\Delta]_{k,s}(\mu_i,m_i)\,, \label{DI.knot} \DIee
with $\zeta=e^{\frac{2\pi i}{k}}$. The prefactor $C=\zeta^{\frac14 f\mb B^{-1}\nu}e^{\frac{i\pi}{4k}(b^2-b^{-2})f\mb B^{-1}\nu}$ depends on an integer solution $(f,f'')\in \BZ^{2N}$ to $\mb A f+\mb B f''=\nu$.

There are several things to note about this partition function:
\begin{itemize}
\item It approximately takes the form of a ``state sum'' or ``state integral,'' with the partition function \eqref{DI.qdl} assigned to every tetrahedron building-block. The ``state variables''  $m_i\in \BZ/k\BZ$ and $\mu_i$ are summed/integrated over. The number of sums/integrals is the same as the number of tetrahedra which is also the same as the number of internal edges in the triangulation.

\item As explained in \cite{DID-levelk,DIDG-levelk}, the partition function is only defined up to an overall phase, of the form $\zeta^{\frac a{24}}e^{\frac{i\pi}{12k}(a'b^2+a''b^{-2})}$, $a,a',a''\in \BZ$. This subsumes the standard \emph{framing ambiguity} in complex Chern-Simons theory \cite{DIWitten-cx}, which would modify the partition function by a factor of $e^{-\frac{i\pi}{2k}(b+b^{-1})^2}$ upon shifting the framing of the tangent bundle of $M$.

\item Each tetrahedron building-block manifestly admits a holomorphic-antiholomorphic factorization as in \eqref{DI.fact}. The full partition function is also expected to admit such a factorization; this was demonstrated for some simple examples in \cite{DID-levelk}. (See also \cite{DIBDP-blocks, DIGK-Gg} for $k=0$, $k=1$ examples.)

\item Perhaps most interestingly, the precise definition of the integration contour in \eqref{DI.knot} and the convergence of the integral depends crucially on the existence of a \emph{positive angle structure} on the triangulation being used. This is ultimately what restricts the computation to a particular class of 3-manifolds that includes all hyperbolic ones. It also beautifully makes contact with the three-dimensional superconformal theory $T[M]$ defined combinatorially in \cite{DIDGG} using an ideal triangulation --- the definition of $T[M]$ does not make sense (cannot produce a superconformal theory) unless a positive angle structure exists. We proceed to explain this idea momentarily.

\item Via the 3d-3d correspondence, the integral \eqref{DI.knot} has a dual interpretation as a partition function of $T[M]$ on the lens space $S^3/\BZ_k$. This partition function is computed by localization methods, as in Section 4.1 of B. Willett's article, \volcite{WI}. Each $\CalZ[\Delta]_{k,s}(\mu_i,m_i)$ is the contribution of a chiral multiplet to the partition function, and the additional prefactor comes from background Chern-Simons terms involving flavor and R-symmetries.

\item Changes of ideal triangulations are generated by 2--3 Pachner moves, which replace a pair of tetrahedra glued along a common face by a triplet glued along three common faces and a central edge. The state-integral \eqref{DI.knot} is invariant under 2--3 moves that preserve the positive or non-negative angle structure (as appropriate), due to a 5-term integral identity for the quantum dilogarithm \eqref{DI.qdl}. In the case $k=1$, this identity was first discovered by Faddeev \cite{DIfaddeev-1994} (see also \cite{DIFad-Kash}).

\end{itemize}

An angle structure on a topological ideal triangulation is an assignment of real parameters (``angles'') to the six long edges of each tetrahedron, in such a way that 1) angles on opposite edges are equal (leaving three angles $\alpha_i,\alpha_i',\alpha_i''$ per tetrahedron $\Delta_i$); 2) the sum of angles around any tetrahedron vertex equals $\pi$ (thus $\alpha_i+\alpha_i'+\alpha_i''=\pi$); and 3) the sum of angles around every internal edge in the triangulation equals $2\pi$. 
In terms of the Neumann-Zagier gluing datum \eqref{DI.NZ} for a knot complement, the last condition translates to%
\footnote{Technically, only $N-1$ out of the $N$ components of \eqref{DI.NZ-angle} correspond to internal edges; the last component of \eqref{DI.NZ-angle} relates to a boundary condition at $\pd M=T^2$, and can be removed from the angle-structure analysis. We are bypassing such subtleties in this heuristic discussion.}
\DIbe \mb A\alpha + \mb B\alpha'' = \pi\nu\,. \label{DI.NZ-angle}\DIee
The idea of an angle structure is motivated by hyperbolic geometry. In an ideal triangulation by \emph{hyperbolic} tetrahedra, the dihedral angles precisely obey the three conditions above \cite{DIthurston-1980}.

\begin{figure}[htb]
\centering
\includegraphics[width=4.8in]{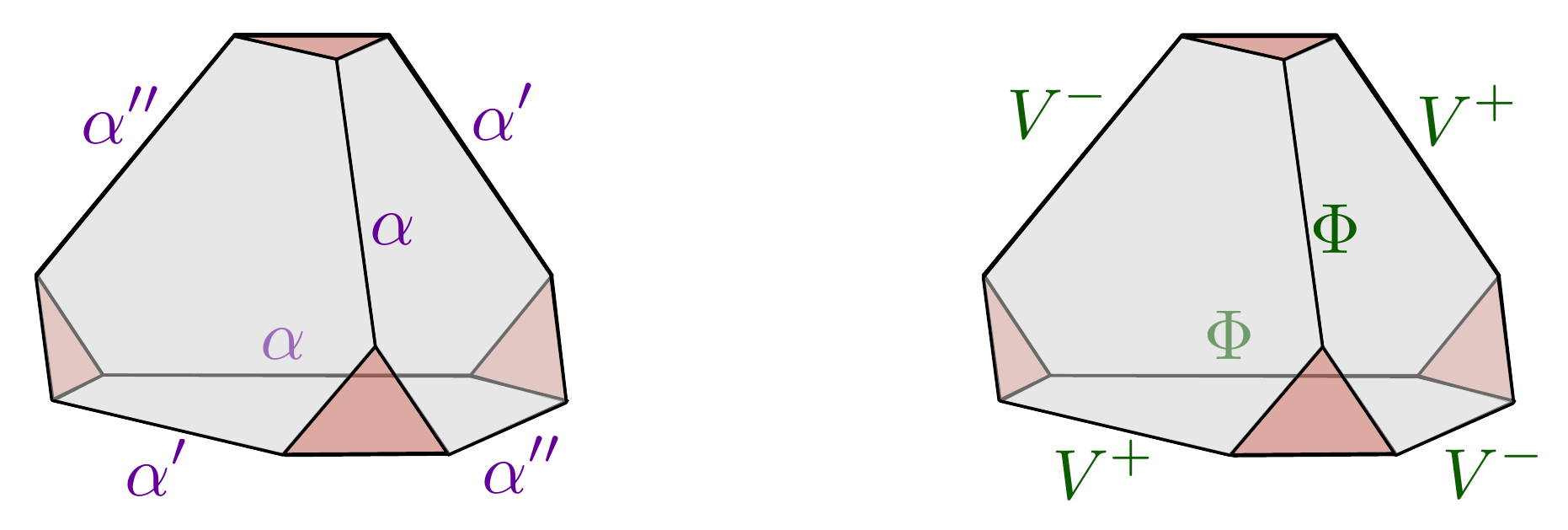}
\caption{Left: dihedral angles assigned to edges of an oriented ideal tetrahedron, obeying $\alpha+\alpha'+\alpha''=\pi$. Right: operators of the tetrahedron theory $T[\Delta]$ assigned to the same edges; these operators have R-charges $\rho,\rho',\rho''$ that obey $\rho+\rho'+\rho''=1$.}
\label{DI.fig:angles}
\end{figure}

A positive (respectively, non-negative) angle structure is one where all angles additionally obey $\alpha_i,\alpha_i',\alpha_i''>0$ ($\geq 0$). An oriented hyperbolic three-manifold with genus-one boundary always admits an triangulation with non-negative dihedral angles, hence there a triangulation with a non-negative angle structure \cite{DIpetronio-2000}. It is conjectured and strongly believed that the same holds for strictly positive angle structures. (As evidence for this, the default triangulations produced by the computer program \texttt{SnapPy} \cite{DISnapPy} for the first few thousand hyperbolic knot complements all admit positive angle structures.)

It was shown in \cite{DIGar-index, DIGHRS-index} that analogue of the partition function \eqref{DI.knot} at $k=0$ (\DIcf\ Section \ref{DI.sec:M2}) is well-defined when the triangulation admits a non-negative angle structure.  Similarly, it was shown in \cite{DIKashAnd} for $k=1$ and \cite{DID-levelk, DIAK-complexCS} for $|k|\geq 1$ that the partition function \eqref{DI.knot} is well defined if the triangulation admits a positive angle structure. In each case, the angle structure specifies a canonical convergent integration contour.

The combinatorial construction of theories $T[M]$ given in \cite{DIDGG} translates the ingredients of the state-integral \eqref{DI.knot} to operations in three-dimensional $\CalN=2$ gauge theories. One starts by associating a theory $T[\Delta_i]$ containing a single chiral multiplet $\Phi_i$ (\DIie\ a free complex boson $\phi_i$ and a complex fermion $\psi_i$) to each tetrahedron.
In addition to the chiral (BPS) operator $\Phi_i$, the theory $T[\Delta_i]$ contains two elementary chiral monopole operators $V_i^\pm$, which exist only in the presence of a monopole background for the flavor symmetry of the theory.%
\footnote{One does not usually talk about ``flavor'' monopole operators like $V_i^\pm$, but they are essential in the construction of theories $T[M]$, and the relation between these theories and geometry. They can easily be seen in the generalized index of a free chiral multiplet \cite{DIKW-index}, and can also be understood as ordinary monopole operators if the flavor symmetry of $T[\Delta_i]$ were weakly gauged.} %
Geometrically, the three operators $\Phi_i,V_i^+,V_i^-$ are associated to three pairs of $\Delta_i$, as on the RHS of Figure~\ref{DI.fig:angles}. Moreover, $T[\Delta_i]$ has an R-symmetry%
\footnote{This is a global symmetry that does not commute with supersymmetry, and is related to an $SO(2)$ automorphism of the 3d $\CalN=2$ SUSY algebra.} %
$U(1)_R$ that rotates $(\phi_i,\psi_i)$ with charges $(\rho_i,\rho_i-1)$ for some (undetermined) $\rho_i\in \BR$. One usually just says that $\Phi_i$ has R-charge $\rho_i$. The other BPS operators $V_i^+,V_i^-$ have some R-charges $\rho_i',\rho_i''$, subject to a single relation $\rho_i+\rho_i'+\rho_i''=1$.

Even for the theory of a single tetrahedron, R-charges are beginning to look like angles, and it is tempting to identify $\rho_i=\alpha_i/\pi$ (etc.). The analogy persists upon gluing. The theory $T[M]$ of \cite{DIDGG} is defined by taking a tensor product of $N$ tetrahedron theories $\otimes_{i=1}^N T[\Delta_i]$, gauging some flavor symmetries, and (most importantly here) adding superpotential interactions $W = \sum_{E=1}^N \CalO_E$ containing a chiral operator $\CalO_E$ for every internal edge $E$ of the triangulation $M=\cup_{i=1}^N \Delta_i$. If all the tetrahedron edges identified with $E$ are labelled by elementary chiral fields $\Phi_i$, one simply constructs $\CalO_E$ as a product $\prod_i\Phi_i$ of these surrounding fields. For edges identified with monopole operators, the prescription is more subtle. In either case, one finds that the R-charge of $\CalO_E$ is a sum of the R-charges $\rho_i,\rho_i',$ or $\rho_i''$ associated to operators on the tetrahedron edges identified with~$E$. The superpotential breaks $U(1)_R$ R-symmetry unless the charge of every $\CalO_E$ is exactly~$2$.

Now, in order for the gauge theory $T[M]$ to flow (in a straightforward way) to a superconformal theory in the infrared, two conditions are necessary: 1) some $U(1)_R$ R-symmetry must be preserved; and 2) there must exist a choice of $U(1)_R$ symmetry such that the charges of all chiral operators are non-negative, because in a superconformal theory R-charges are proportional to operator dimensions. This means that we must be able to choose $\rho_i,\rho_i',\rho_i''$ for individual tetrahedra in such a way that the sums of these charges around any internal edge $E$ is 2, and all these charges are non-negative. Together with the relation $\rho_i+\rho_i'+\rho_i''=1$, the conditions become equivalent to the existence of a non-negative angle structure.

\subsection{Embedded surfaces and M2 branes in the 3d index ($k=0$)}
\label{DI.sec:M2}

The partition function of complex Chern-Simons theory at level $k=0$ is rather special. Via the 3d-3d correspondence, it takes the form of a supersymmetric index \cite{DIDGG-index}. Schematically:
\DIbe \CalZ[M]_{k=0,s} = \CalZ[\text{$T[M]$ on $S^2\!\times_q\!S^1$}] = \text{Tr}_{\CalH[\text{$T[M]$ on $S^2$}]} (-1)^Rq^{\frac{R}{2}+J_3}\,, \label{DI.index} \DIee
with $q=\exp\frac{4\pi}{2}$ as usual.
Here $S^2\!\times_q\!S^1$ denotes a geometry that is fibered over $S^1$, in such a way that $S^2$ rotates around an axis (by an amount $\pi/s$) as $S^1$ is traversed. Just as in our discussion around \eqref{DI.tr}, the partition function of any quantum field theory in such a geometry can be expressed as a trace. In this case, it becomes a trace over the Hilbert space of $T[M]$ on $S^2$, weighted by the spin ($J_3$) and R-charge ($R$) of states. Due to the factor $(-1)^R$, this trace behaves like an Euler character (physically, a Witten index \cite{DIWitten-constraints}). In particular, only ``BPS states'' in the cohomology of a particular supercharge contribute. While the ``BPS sector'' of the Hilbert space containing such states is infinite-dimensional, it is expected to have finite $q$-graded dimension when $M$ is sufficiently nice (\DIeg\ hyperbolic), so that \eqref{DI.index} produces a well-defined formal Laurent series in $q$.
Less obviously, if $T[M]$ is a superconformal theory, then superconformal symmetry requires the existence of an R-charge assignment such that only non-negative powers of $q$ appear in \eqref{DI.index} \cite{DIDGG-index}.

In the case that $T[M]$ is superconformal, one can also use the state-operator correspondence to recast the RHS or \eqref{DI.index} as a sum over BPS \emph{operators} of $T[M]$ rather than states.
Then the index counts the number of operators with given spin and R-charge. 
Combining this perspective with the 3d-3d correspondence really makes the geometry of $M$ come to life, in the following way.

Recall that for gauge group $G_\BC=SL(N,\BC)$, the theory $T[M]$ is the effective theory of $N$ M5 branes wrapped on $M\times \BR^3$ inside the eleven-dimensional background $T^*M\times \BR^5$. From this perspective, at least some of the BPS operators in $T[M]$ should come from M2 branes that end on the stack of M5 branes, such that their boundary $\pd$(M2)$=\Sigma\times\{0\} \subset M\times \BR^3$ wraps a surface in $M$. One might therefore expect that
\DIbe \text{The index $\CalZ[M]_{k=0,s}$ ``counts'' surfaces in $M$.} \label{DI.counts} \DIee

Alternatively, we may think of $T[M]$ as the effective theory obtained by compactifying the 6d $(2,0)$ theory of type $A_{N-1}$ on $M$. The local BPS operators of $T[M]$ can come from 1) local operators of the 6d theory; or 2) surface operators in the 6d theory, corresponding to the boundaries of M2 branes in the M-theory picture above. Thus, a more refined statement of \eqref{DI.counts} would be that the index counts surfaces in $M$ decorated by local operators.

The first hint that \eqref{DI.counts} could be true came in work of Gukov, Gadde, and Putrov \cite{DIGGP-walls}. They investigated the index for knot complements $M=S^3\backslash K$, and found contributions from boundary-incompressible surfaces. These are incompressible surfaces $\Sigma$ whose boundary $\pd \Sigma = S^1\subset \pd M$ wraps a particular nontrivial cycle on the torus boundary of $M$. The corresponding operators $\CalO_\Sigma$ and all of their powers $\CalO_\Sigma^2$, $\CalO_\Sigma^3$,...  (corresponding to multiply-wrapped surfaces) are expected to contribute to the index, and this is exactly what \cite{DIGGP-walls} found.

To make a more precise statement, we should recall that the index $\CalZ[M]_{k=0,s}(m,n)$ of a knot complement  depends on two integers $m,n$, as in \eqref{DI.index-fmn}. On one hand, these integers are magnetic and electric (flavor) charges in the space of local BPS operators of $T[M]$. On the other hand, these charges label homology classes $(m,n)\in H_1(\pd M,\BZ)$.
For an operator $\CalO_\Sigma$ corresponding to a surface in $M$, the charges $(m_\Sigma,n_\Sigma)$ measure the homology class of the boundary curve $\pd \Sigma\subset \pd M$.

In the case of a boundary-incompressible surface $\Sigma$, the expectation is that the operator $\CalO_\Sigma$  preserves not one but two supercharges. This implies that if $\CalO_\Sigma$ has $\frac R2+J_3=Q_\Sigma$, so that it contributes $q^{Q_\Sigma}$ to the index in some charge sector $(m_\Sigma,n_\Sigma)$, then the $d$-th power $\CalO_\Sigma^d$ has $\frac R2+J_3=dQ_\Sigma$ and contributes $q^{dQ_\Sigma}$ to the index in charge sector $(dm_\Sigma,dn_\Sigma)$. One can therefore detect the presence of such operators in the index because they force the minimal power of $q$ appearing in $\CalZ[M]_{k=0,s}(dm,dn)$ to grow linearly as $d\to \infty$ whenever $(m,n)$ coincides with the charge $(m_\Sigma,n_\Sigma)$ of an incompressible surface.
In contrast, for generic $(m,n)$, the growth of $\CalZ[M]_{k=0,s}(dm,dn)$ as $d\to \infty$ is quadratic.%
\footnote{Similar behavior appears in certain stable limits of the Jones polynomial \cite{DIGaroufalidis-slopes}. In both cases, the behavior can partially be explained by observing that 1) one is studying $q$-series that are solutions to recursion relations coming from the $A$-polynomial of a knot \cite{DIgukov-2003, DIgaroufalidis-2004, DIDGG-index}; and 2) boundary slopes of the $A$-polynomial are related to boundary slopes of incompressible surfaces \cite{DIcooper-1994}. See \cite[App. D]{DIDGG-index} for further remarks.} %
Famously, the set of possible charges of incompressible surfaces $(m_\Sigma,n_\Sigma)$ is finite \cite{DIHatcher-bdy}. 

\begin{figure}[htb]
\centering
\includegraphics[width=5in]{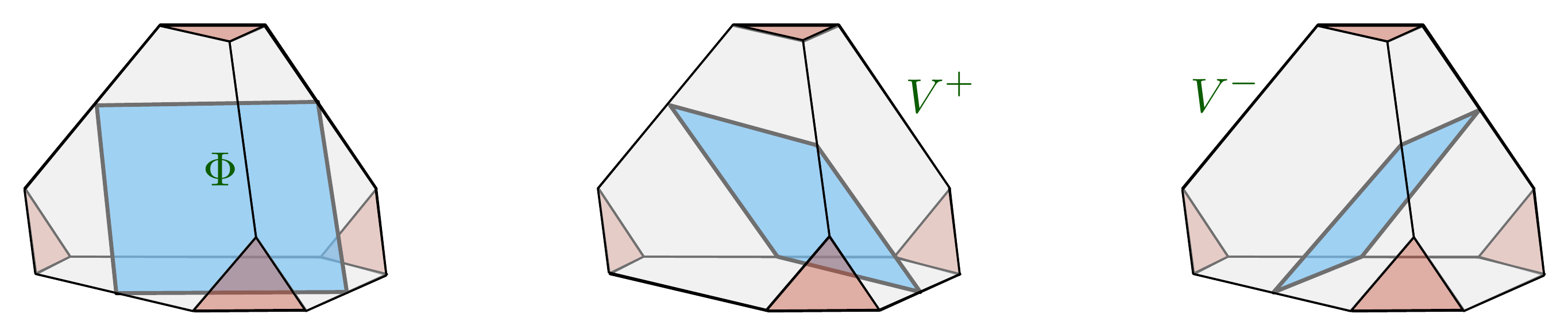}
\caption{Three incompressible surfaces inside an ideal tetrahedron, corresponding to three unconstrained operators in the theory $T[\Delta]$.}
\label{DI.fig:surface}
\end{figure}

The simplest example of such behavior occurs not for a knot complement but for a single ideal tetrahedron $M=\Delta$. The index $\CalZ[\Delta]_{k=0,s}(m,n)$ is defined by
\DIbe \sum_{n\in \BZ} \CalZ[\Delta]_{k=0,s}(m,n)\, \zeta^n = \prod_{r=0}^\infty \frac{1-q^{1+r}x^{-1}}{1-q^r \tilde x^{-1}}\,, \label{DI.tet-index} \DIee
where $x=q^{\frac m2}\zeta$ and $\tilde x=q^{\frac m2}\zeta^{-1}$ (\DIcf\ \eqref{DI.index-fmz}). Explicitly,
\DIbe \CalZ[\Delta]_{k=0,s}(m,n) = \sum_{r=(-n)_+}^\infty \frac{q^{-\big(r+\frac12n\big)m}}{\prod_{i=1}^r(1-q^{-i})\prod_{j=1}^{r+n}(1-q^j)}\,, \DIee
where $(-n)_+=\text{max}(-n,0)$.
There are three rays in the $(m,n)$ lattice along which the leading power of $q$ grows linearly: $\gamma=(0,-d)$, $\gamma'=(d,0)$, and $\gamma''=(-d,d)$ for $d\geq 0$. In terms of the theory $T[\Delta]$, these three rays correspond to powers of the three operators $\Phi,V^+,V^-$ that were discussed at the end of Section \ref{DI.sec:angle}. Geometrically, they correspond to the three ``local'' incompressible surfaces shown in Figure \ref{DI.fig:surface}. (These surfaces are boundary-incompressible if the boundary of the ideal tetrahedron is understood as a four-punctured sphere. They are the basic building blocks of what are known as normal surfaces in three-manifold topology~\cite{DIRubinstein-S3}.)

The boundary-incompressible surfaces in $M$ lead to very special operators that are unconstrained, in the sense that all their powers contribute to the index, with purely additive $\frac R2+J_3$ charge.
In general, one might expect to find operators whose powers do obey various relations, and do not have additive charges.  Excitingly, in recent mathematical work \cite{DIGHHR}, 
 Hodgson, Hoffman, Garoufalidis, and Rubinstein found that it was possible to rewrite the \emph{entire} index 
 $\CalZ[M]_{k=0,s}$ of a knot complement as a sum over surfaces,
 \DIbe \CalZ[M]_{k=0,s}(m,n)\; = \hspace{-.2cm}\sum_{\text{$\Sigma$ s.t. $[\pd \Sigma]=(m,n)$}} \hspace{-.2cm}
 w_\Sigma(q)\,, \DIee
 where the sum is over all normal surfaces in a given triangulation of $M$, making the expected relation \eqref{DI.counts} true in a precise sense.
 The main contribution to the weights $w_\Sigma(q)$ looks like a product of tetrahedron indices, corresponding to the tetrahedra in the triangulation. It would be very interesting to understand these weights in terms of local operators of the 6d $(2,0)$ theory, bound to the surface $\Sigma$.

\section{The Quantum Modularity Conjecture}
\label{DI.sec:QM}

Our final topic concerns the relation between complex Chern-Simons theory and Zagier's Quantum Modularity Conjecture (QMC) \cite{DIQMF}. Strictly speaking, the QMC concerns certain limits of colored Jones polynomials of knots in $S^3$. Physically, colored Jones polynomials have to do with $SU(2)$ Chern-Simons theory. However, by embedding $SU(2)$ theory into a holomorphic sector of $SL(2,\BC)$ theory, one can use complex Chern-Simons theory to study the very same limits \cite{DIDG-levelk}. Moreover, $SL(2,\BC)$ theory and its various stringy realizations may have a chance of explaining some of the subtler aspects of the conjecture.

To formulate a precise statement, we begin by recalling that the colored Jones polynomials $J_K(N,q)$ of a knot $K\subset S^3$ are expectation values in $SU(2)$ Chern-Simons theory of Wilson loops in the $N$-dimensional representation \cite{DIJones, DITuraev-YB, DIWit-Jones}. Physically, the parameter $q=\exp\tfrac{2\pi i}{k}$ encodes the renormalized level $k\in \BZ$ of the $SU(2)$ theory, but (as the name suggest) the Jones polynomials can be analytically continued to Laurent polynomials in a formal variable~$q$. For any $k$-th root of unity $q^k=1$, the Jones polynomials are periodic $J_K(N,q)=J_K(N+k,q)$.

The colored Jones polynomials can equivalently be defined using the wavefunction of $SU(2)$ Chern-Simons at renormalized level $k$ on the knot complement $M=S^3\bs K$. (The bare level appearing in the Lagrangian is $k-2$.) Recall, \DIeg\ from Section \ref{DI.sec:quant}, that the Hilbert space $\CalH[T^2]_k$ in compact Chern-Simons theory is $k-1$ dimensional. The $SU(2)$ wavefunction $\CalZ[M]_k$ is thus a vector in $\BC^{k-1}$, whose components are precisely the Jones polynomials $J_K(1,e^{\frac{2\pi i}{k}}),\,J_K(2,e^{\frac{2\pi i}{k}}),...,\,J_K(k-1,e^{\frac{2\pi i}{k}})$.

For any rational number $\alpha=a/c$, Zagier defines
\DIbe \CalJ_K(\alpha) := J_K(c,e^{\frac{2\pi ia}{c}})\,. \label{DI.Ja} \DIee
Crucially, this is just outside the natural range of states in $SU(2)$ Chern-Simons at renormalized level $c$, hinting that some analytic continuation will be necessary for a physical interpretation of \eqref{DI.Ja} (just like it was in the Volume Conjecture).

The QMF states in part that $\CalJ_K(\alpha)$ shows exponential growth as $\alpha$ tends to any fixed rational number $\alpha_0$ through rational values, with the rate of growth governed by the hyperbolic volume of $M=S^3\bs K$. Specifically, let $\left(\begin{smallmatrix} a&b\\c&d\end{smallmatrix}\right)\in SL(2,\BZ)$ with $c>0$, and let $X\to\infty$ in a fixed coset of $\BQ/\BZ$  (for example: $X=1,2,4,5,...$ or $X=\frac34,\frac74,\frac{11}4,...$ with constant denominators). Set $\hbar= 2\pi i/(cX+d)$. Then, conjecturally, there is an asymptotic expansion
\DIbe \CalJ_K\left(\frac{aX+b}{cX+d}\right) \sim \CalJ_K(X) \left(\frac{2\pi i}{\hbar}\right)^{\frac32} e^{\tfrac{1}{c\hbar}S_{CS}(\text{geom})} \phi_{K,a/c}(\hbar)\,, \label{DI.QMC} \DIee
where $S_{CS}(\text{geom}) = \text{CS}(M)+i\text{Vol}(M)$ is the volume of the geometric flat connection on $M$, containing the hyperbolic volume as in Section \ref{DI.sec:vol}; and  $\phi_{K,a/c}(\hbar)$ is a formal power series in $\hbar$ that depends only on $a/c$ mod $1$. Moreover, it was conjectured that after dividing by the leading coefficient to make $\widetilde \phi_{K,a/c}(\hbar) =  \phi_{K,a/c}(\hbar)/ \phi_{K,a/c}(0)$ monic, the subleading coefficients all belong to the trace field of $K$ adjoined a $c$-th root of unity $\Gamma_K(e^{\frac{2\pi i a}{c}})$. (The number field $\Gamma_K$ contains the traces of all holonomies of the geometric flat $SL(2,\BC)$ connection on $M$.)

For $\left(\begin{smallmatrix} a&b\\c&d\end{smallmatrix}\right) = \left(\begin{smallmatrix} 0&-1\\1&0\end{smallmatrix}\right)$ and $X=1,2,3,...\in \mathbb N$, the QMC \eqref{DI.QMC} reduces to the usual Volume Conjecture.%
\footnote{The number-theoretic properties of the asymptotic expansion in the Volume Conjecture were explored from a physical perspective in \cite{DIDGLZ}, leading to the notion of an Arithmetic Quantum Field Theory.} %
Heuristically, one may think of the Volume Conjecture as studying a limit of colored Jones polynomials as $q\to 1$, and the QMC as studying limits where $q$ approaches all the rational points on the unit circle.

In order to embed the QMC in $SL(2,\BC)$ Chern-Simons theory, one can consider singular limits very similar to \eqref{DI.sing1}.%
\footnote{There may be other ways to embed the QMC into complex Chern-Simons theory. It was proposed in \cite{DIGK-rational} that setting $(k+is,k-is)\to(a,c)$ to finite integer values, which is a perfectly regular limit in complex Chern-Simons, may also be related to the QMC.} %
 Namely, it was argued in \cite{DIDG-levelk} that when fixing the integer level $k$ and sending $s\to -ik$ in complex Chern-Simons theory, or equivalently $(k+is,k-is)\to(2k,0)$, the knot-complement partition function has an asymptotic expansion
\DIbe\qquad \CalZ[M]_{k,s} \overset{s\to -ik}\sim \left(\frac{2\pi i}{\hbar}\right)^{\frac32} e^{\tfrac{1}{k\hbar}S_{CS}(\text{geom})} \phi_{K,1/k}(\hbar)\,,\qquad \hbar = 2\pi i\frac{k-is}{k+is} \label{DI.QMC-cx} \DIee
with the same series as in \eqref{DI.QMC}. Using the state-sum construction \eqref{DI.knot} (on one hand) and detailed computations of Garoufalidis and Zagier for asymptotics of colored Jones polynomials (on the other hand), the agreement between \eqref{DI.QMC} and \eqref{DI.QMC-cx} was verified in dozens of examples. By performing a saddle-point expansion of the state-sum model, one also deduces that coefficients of $\widetilde\phi_{K,1/k}(\hbar)$ lie in the number field $\Gamma_K(e^{\frac{2\pi i}{k}})$ as desired.

In terms of the 3d-3d correspondence, the singular limit in \eqref{DI.QMC-cx} corresponds to computing the partition function of $T[M]$ on a squashed lens space $L(k,1)_b \simeq S^3_b/\BZ_k$, and sending to zero the squashing parameter $b^2\to 0$. The $1/k$ dependence in the leading asymptotic can be explained from the simple fact that the lens space is a $k$-fold quotient of the three-sphere. It would be interesting to make further physical predictions for the QMC by relating the partition functions of $T[M]$ on different lens spaces $L(k,1)_b$, or more generally $L(c,a)_b$.

\begin{figure}[htb]
\centering
\includegraphics[width=4in]{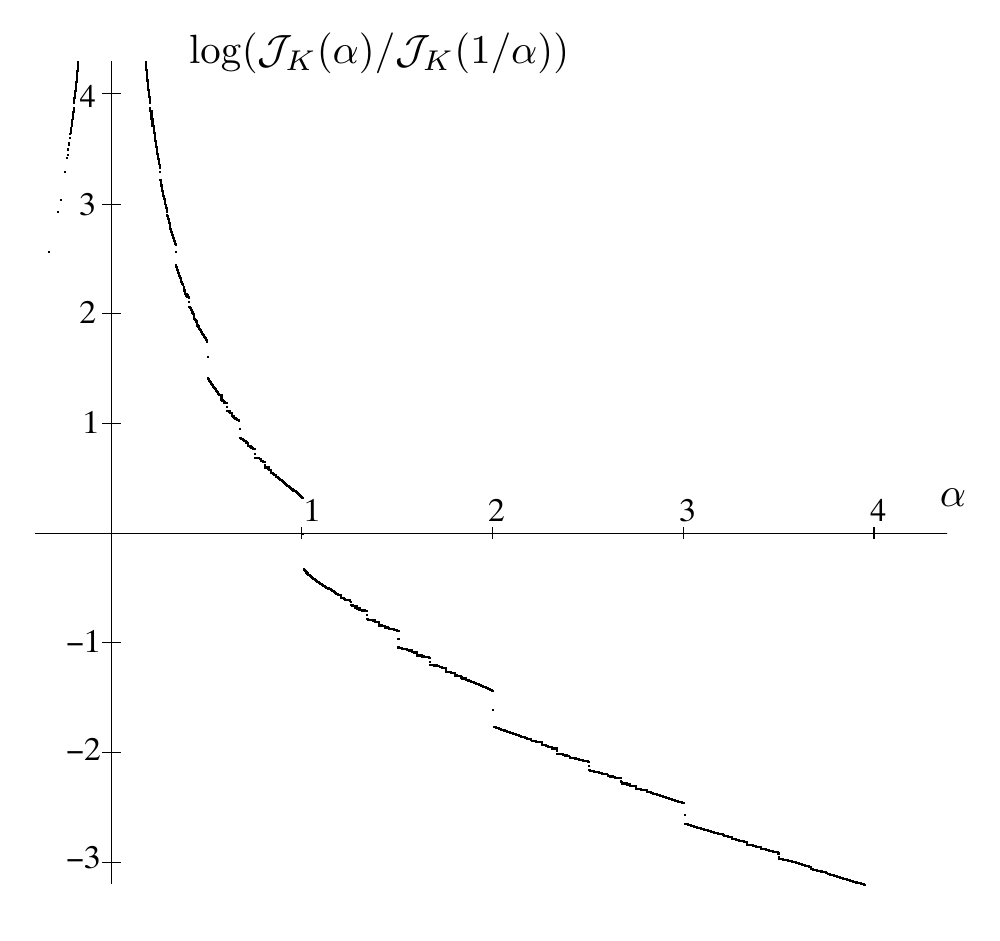}
\caption{The graph of $\CalJ_K(\alpha)/\CalJ_K(1/\alpha)$ when $K$ is the figure-eight knot, from \cite{DIQMF}.}
\label{DI.fig:QMF}
\end{figure}

Another observation of \cite{DIQMF} is that, for hyperbolic several knots, the ratio $\CalJ_K(\alpha)/\CalJ_K(\gamma(\alpha))$ for $\gamma\in SL(2,\BZ)$ has some very interesting finite behavior. This ratio measures the \emph{failure} of $\CalJ_K(\alpha)$ to be truly ``modular.'' For the figure-eight knot, a graph of this function appears in \cite[Fig. 4]{DIQMF}, reproduced above. The graph is almost monotonic (it violates monotonicity at a very fine scales) and discontinuous at all rational points $a/c$, the size of the discontinuity being roughly proportional to the denominator $|c|$.

If we turn this graph on its side, any condensed matter theorist would recognize it immediately: it is the plot of Hall resistivity as a function of magnetic field in a two-dimensional electron system. The plot has plateaus at rational points, due to the fractional quantum Hall effect. It is extremely tempting to think that this must be related to the effective complex Chern-Simons theory in quantum-Hall systems \cite{DIVafa-FQHE}; the connection is investigated in \cite{DIDGV-FQHE}.


\subsection*{Acknowledgements}

I am grateful to Stavros Garoufalidis, Sergei Gukov, Neil Hoffman, Craig Hodgson, Hyam Rubinstein, Daniel Jafferis, and Cumrun Vafa for many insightful discussions on the themes that entered this review.

\documentfinish